\documentclass[sigconf]{acmart}
\AtBeginDocument{%
  \providecommand\BibTeX{{%
    \normalfont B\kern-0.5em{\scshape i\kern-0.25em b}\kern-0.8em\TeX}}}

\copyrightyear{2021}
\acmYear{2021}
\setcopyright{acmcopyright}
\acmConference[ICTIR '21] {Proceedings of the 2021 ACM SIGIR International Conference on the Theory of Information Retrieval}{July 26, 2021}{Virtual Event, Canada.}
\acmBooktitle{Proceedings of the 2021 ACM SIGIR International Conference on the Theory of Information Retrieval (ICTIR '21), July 26, 2021, Virtual Event, Canada}
\acmPrice{15.00}
\acmDOI{10.1145/3471158.3472245}
\acmISBN{978-1-4503-8611-1/21/07}
\definecolor{midnightgreen}{rgb}{0.0, 0.29, 0.33}
\definecolor{darkpink}{rgb}{0.91, 0.33, 0.5}

\usepackage{multirow}
\usepackage{subcaption}
\usepackage{balance}

\begin{document}
% \title{Dense Retrieval with Contrastive Dual Learning}
\title{More Robust Dense Retrieval with Contrastive Dual Learning}
\fancyhead{}

\author{Yizhi Li$^{1*}$, Zhenghao Liu$^{1*}$, Chenyan Xiong$^2$, and Zhiyuan Liu$^1$}\thanks{$*$\ \ indicates equal contribution.}
\affiliation{Tsinghua University$^1$, Microsoft Research$^2$ 
\country{}
} 
\affiliation{
\texttt{yizhi.li@hotmail.com};
\texttt{liuzhenghao0819@gmail.com};
\texttt{chenyan.xiong@microsoft.com};
\texttt{liuzy@tsinghua.edu.cn}
\country{}
}
\begin{abstract}
Dense retrieval conducts text retrieval in the embedding space and has shown many advantages compared to sparse retrieval. Existing dense retrievers optimize representations of queries and documents with contrastive training and map them to the embedding space. The embedding space is optimized by aligning the matched query-document pairs and pushing the negative documents away from the query. However, in such training paradigm, the queries are only optimized to align to the documents and are coarsely positioned, leading to an anisotropic query embedding space. In this paper, we analyze the embedding space distributions and propose an effective training paradigm, Contrastive Dual Learning for Approximate Nearest Neighbor (\texttt{DANCE}) to learn fine-grained query representations for dense retrieval. \texttt{DANCE} incorporates an additional dual training object of query retrieval, inspired by the classic information retrieval training axiom, query likelihood. With contrastive learning, the dual training object of \texttt{DANCE} learns more tailored representations for queries and documents to keep the embedding space smooth and uniform, thriving on the ranking performance of \texttt{DANCE} on the MS MARCO document retrieval task. Different from \texttt{ANCE} that only optimized with the document retrieval task, \texttt{DANCE} concentrates the query embeddings closer to document representations
while making the document distribution more discriminative. Such concentrated query embedding distribution assigns more uniform negative sampling probabilities to queries and helps to sufficiently optimize query representations in the query retrieval task. Our codes are released at \url{https://github.com/thunlp/DANCE}.
\end{abstract}




\maketitle

\section{Introduction}
The recent dense retrieval provides an opportunity to conduct semantic matches and serves lots of applications, such as open domain question answering~\cite{chen2017reading}, conversational search~\cite{qu2020orqa,Yu2021FewShotCD}, and fact verification~\cite{thorne2018fact}. Instead of using discrete bag-of-word matching in document retrieval, dense retriever encodes queries and documents into a high-dimensional embedding space and conducts text matching in the embedding space, overcoming the vocabulary mismatch problem of sparse retrieval.

Existing dense retrieval models aim to learn effective representations of queries and documents and build an embedding space for the document retrieval task. With the support of approximate nearest neighbor (ANN) search~\cite{johnson2019billion, guo2020accelerating}, dense retrievers can efficiently retrieve documents by conducting semantic matching in the embedding space. To encode queries and documents as dense representations, dense retrievers usually employ the BERT-Siamese architecture to provide fully learnable, well pretrained, and effective representations for queries and documents to construct embedding space for retrieval~\cite{karpukhin2020dense,xiong2020dense,xiong2020approximate}.

To optimize the embedding space for document retrieval, existing dense retrieval models usually sample negative documents and use them to contrastively train dense retrievers to learn query and document representations~\cite{xiong2020approximate,xiong2020dense,karpukhin2020dense}. Wang et al.~\cite{wang2020understanding} prove that contrastive representation learning optimizes neural models and keeps two properties of the document embedding space, ``alignment'' and ``uniformity''. The two properties, ``alignment'' and ``uniformity'', help to keep a homogeneous and isotropic embedding space~\cite{xiong2020approximate,xiong2020dense,karpukhin2020dense} -- which is that ``alignment'' assigns similar embedding features to query and its related documents and conducts better clustering for similar document representations; ``uniformity'' encourages encoders to maintain maximal information for documents. It helps to form a uniform document embedding space and better classifies the confusable documents according to queries.
% ``uniformity'' encourages encoders to maintain maximal information for documents, discards unnecessary details in learning the document representations, and helps to form a uniform document embedding space to better classify the confusable documents according to queries. 
The training object in existing dense retrieval methods~\cite{xiong2020approximate,xiong2020dense,karpukhin2020dense} is to train dense retrievers on the document retrieval task and focuses on optimizing document embeddings. However, such document retrieval dominated training paradigm only optimizes the queries aligned to the relevant documents while discarding the uniformity nature in contrastive learning, leading to a non-smooth and non-uniform query embedding space.
Thus, some work~\cite{zhan2020learning} solely optimizes the query representations by training with full retrieved documents
instead of negative sampled documents and fixing the document representations.

In this paper, we enhance the query representations learning with a new training paradigm, Contrastive Dual Learning for Approximate Nearest Neighbor (\texttt{DANCE}). \texttt{DANCE} uses a model-level dual training object~\cite{xia2018model} to contrastively train BERT for optimizing the query and document embedding spaces. It consists of two optimization directions, document retrieval and query retrieval, which are the main training task and dual training task, respectively. 
% The query retrieval task requires the model to retrieve related queries with given documents, which is in the reverse direction of the document retrieval task. 
% The query retrieval task .
Query retrieval learns the query likelihood~\cite{lavrenko2017lm} to retrieve related queries for documents, and mainly focuses on optimizing query representations in consideration of the document representations near to them. 
% Query retrieval learns the query likelihood~\cite{lavrenko2017lm}, and mainly focuses on optimizing query representations in consideration of the document representations near to them. 
Intuitively, \texttt{DANCE} can learn more appropriate query representations and better keep ``alignment'' and ``uniformity'' of the query embedding space, thus improving the robustness of the learned representation space. Besides, \texttt{DANCE} also normalizes the representations to avoid overfitting during training~\cite{chen2020simple}. We map embeddings into a unit hyperspherical space and calculate the similarity between embeddings according to their angles.

% These well trained nearest query embeddings help to assign reasonable positions for the long tailed documents in the embedding space. Intuitively, DANCE can give more appropriate representations for long tailed documents, better keep ``alignment'' and ``uniformatity'' among embeddings and achieve a more smooth, aligned and uniform embedding space for better retrieval.

%Inspired by the train axiom traditional statistical language model~\cite{lavrenko2017lm} in IR, the query likelihood can also be used to approximate the relevance between query and document according to the Bayesian theorem, showing effectiveness in improve ranking accuracy for both traditional and neural IR models~\cite{nogueira2019doc2query,ma2020prop}.
%It assumes that the related queries can be generated with a given document and teaches models to distinguish different queries. The query likelihood provides an opportunity to train dense retriever to generate more discriminative representations for queries.
%DANCE provides an additional regularization (query likelihood) to fully use the relevance signals to optimize the pre-trained language models and guide the dense retriever to also focus on distinguishing different queries according to the given document.

Experiments on the MS MARCO~\cite{bajaj2016ms} document retrieval task show the effectiveness of \texttt{DANCE}. \texttt{DANCE} mainly focuses on optimizing the query representations by training dense retrievers with the additional query retrieval task. It helps to build a smooth and uniform embedding space for text retrieval by pushing the query embeddings closer to document embeddings. Moreover, \texttt{DANCE} scatters the document distribution to make them more discriminative. Our analyses find that the queries are assigned with more balanced and uniform probabilities to be recalled in the contrastive training of query retrieval task, which sufficiently trains the query representations. For the documents recalled more easily in the contrastive training, they are usually distributed in a more concentrated area in the embedding space, which are confusable. \texttt{DANCE} shows better performance on these documents by learning more fine-grained query representations, which helps to distinguish the off-topic and unrelated documents during retrieval.

%In the rest of this paper, Section 2 discusses related work and Section 3 presents our \texttt{DANCE} method.
%Experiment settings and evaluation results are discussed in Section 4 and Section 5, respectively.
%We conclude in Section 6.
\section{Related Work}
Based on whether term-level interactions are modeled between query and documents beyond their final encodings, Neural IR (Neu-IR) methods can be categorized into representation-based or interaction-based~\cite{jiafeng2016deep, mitra2018introduction}.

Interaction-based models enjoy fine-grained modeling of term-level interactions between query and documents; thus they are typically more effective though more expensive and usually used as re-rankers since that requires scoring candidate documents according to the given query~\cite{xiong2017knrm, convknrm, nogueira2019passage, jiafeng2016deep, macavaney2019cedr, hui2017pacrr}. 
Representation-based ones, often encode queries and documents as low-dimensional dense representations without explicit term-level matches. The representation-based models can achieve more efficient retrieval with the document representation precomputing and the support of approximate nearest neighbor (ANN)~\cite{xiong2020approximate, karpukhin2020dense, lee2019latent, luan2020sparse, gao2020complementing, khattab2020colbert}. The representation-based models help to achieve an efficient dense retrieval, benefiting many downstream tasks by providing more accurate evidence, such as fact verification, conversational dense retrieval, and open domain question answering~\cite{lewis2020rag,karpukhin2020dense,xiong2020dense,qu2020orqa}.

With the development of the deep neural network, pretrained language models, such as BERT~\cite{liu2019roberta} and RoBERTa~\cite{devlin2019bert}, have been widely used in both representation-based and interaction-based Neu-IR models.
Dense retrievers also employ the BERT-Siamese architecture to encode queries and documents as dense representations to conduct an embedding space for retrieval. To learn query and document representations, dense retrievers contrastively train BERT on the document retrieval task with related (positive) document and sampled unrelated (negative) documents for the given queries. 
While some work~\cite{khattab2020colbert} proposes a late query–document interaction paradigm to improve the retrieval effectiveness, many others focus on learning better representations with contrastive training. They propose various negative sampling methods during contrastive training, including in-batch negatives, BM25 retrieved negatives and random selected negatives~\cite{xiong2020dense,xiong2020approximate,lewis2020pre}.
% The negative documents are crucial to learning query and document representations in the embedding space. Some existing work fine-tunes the retrievers with negative documents from in-batch documents, BM25 retrieved documents and random selected documents~\cite{xiong2020dense,xiong2020approximate,lewis2020pre,khattab2020colbert}. 
Approximate Nearest Neighbor Negative Contrastive Learning (\texttt{ANCE})~\cite{xiong2020approximate} further comes up with a training method that asynchronously updates the document index and samples negatives from queries' nearest areas in the document embedding space to avoid diminishing gradient norms during training~\cite{xiong2020approximate}. Some work~\cite{prakash2021LearningRobustDense} also improves such training paradigm by alleviating the affect from incomplete relevant labels.

Recent work in computer vision further discusses how the embedding space is optimized with contrastive training~\cite{chen2020simple}. The contrastive learning for dense retrievers optimizes the embedding space to satisfy two properties, ``alignment'' and ``uniformity''~\cite{wang2020understanding}, which aim to align matched query-document pairs and make the embedding space uniform by pushing the negative documents away from the given queries, respectively.
Several work demonstrates that mapping the representations into a unit hyperspherical space, where all embeddings are represented as unit vectors, helps to keep a smooth embedding space and brings improvement for various tasks~\cite{liu2017sphereface,wang2017normface,wang2020understanding,chen2020simple,chen2020intriguing}. Chen et al.~\cite{chen2020simple} introduce the normalized temperature-scaled cross entropy loss as the standard contrastive training loss. Different from the standard cross entropy loss, it uses the temperature scaling technique to adjust the sharpness of the softmax distribution, which plays an important role in optimizing the normalized embedding space~\cite{chen2020simple}. Moreover, Chen et al.~\cite{chen2020intriguing} further adjust the temperature and balance the influence of ``alignment'' and ``uniformity'' to learn a better embedding space with contrastive training.

\begin{figure*}[t]
    \centering
    \setlength{\lineskip}{\medskipamount}
    
    \subcaptionbox{
        Embedding Space of \texttt{ANCE}. \label{fig:loss_insight_a}
    }{
        \includegraphics[width=0.215\textwidth]{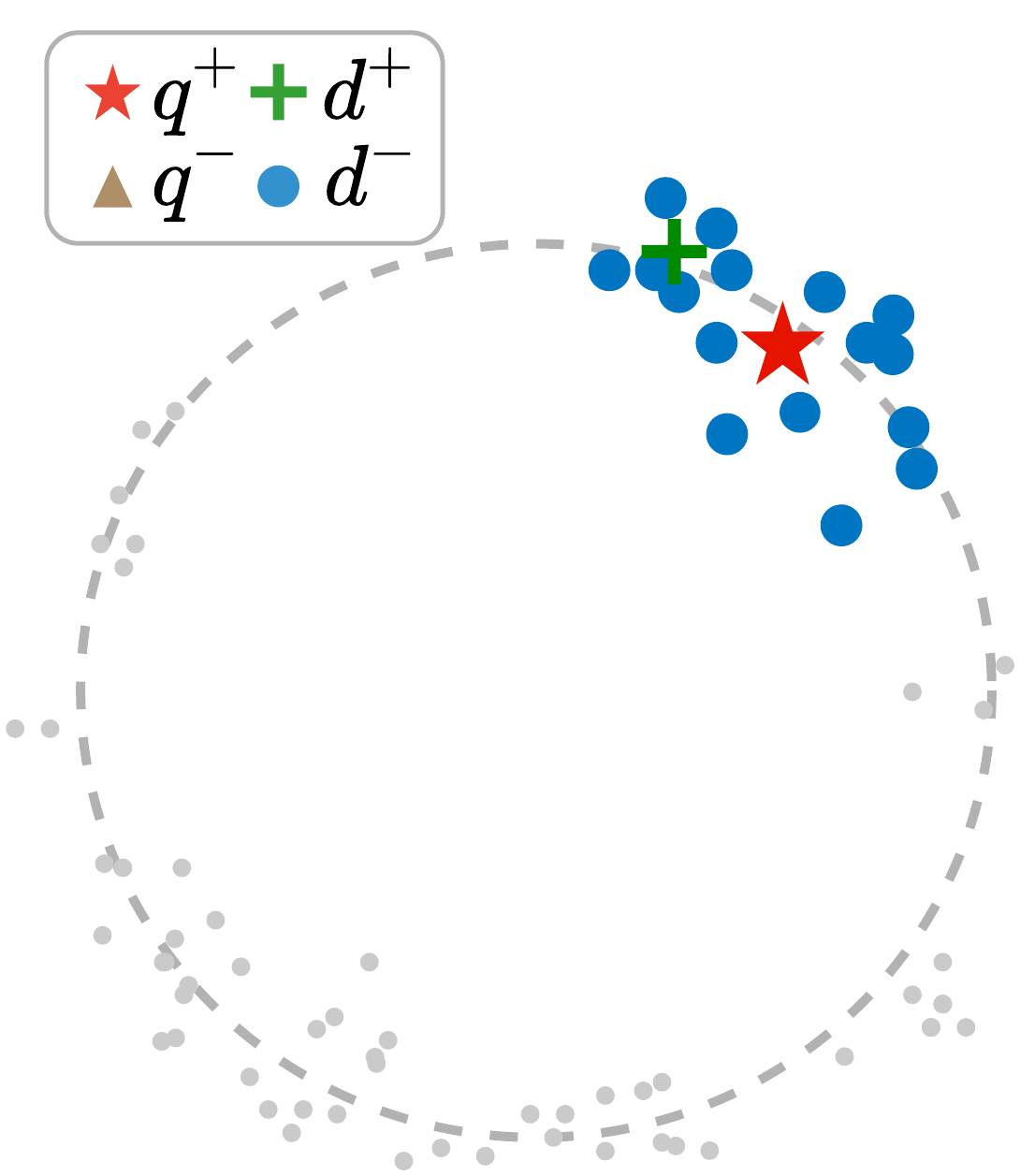}
    } %\hfill    
    \subcaptionbox{
        Document Retrieval (Main). \label{fig:loss_insight_b}
    }{
        \includegraphics[width=0.215\textwidth]{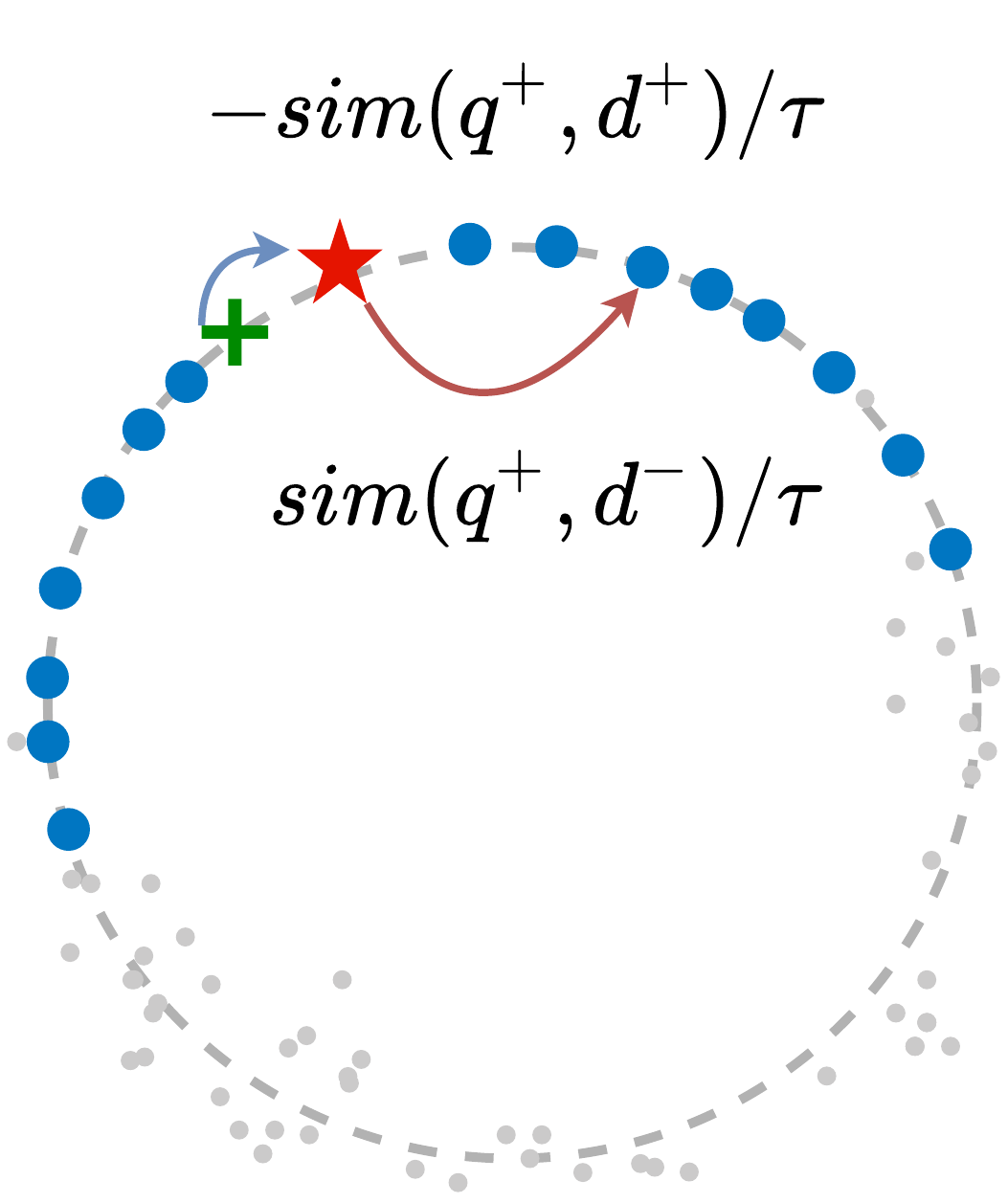}
    } %\hfill    
    \subcaptionbox{
        Query Retrieval (Dual). \label{fig:loss_insight_c}
    }{
        \includegraphics[width=0.215\textwidth]{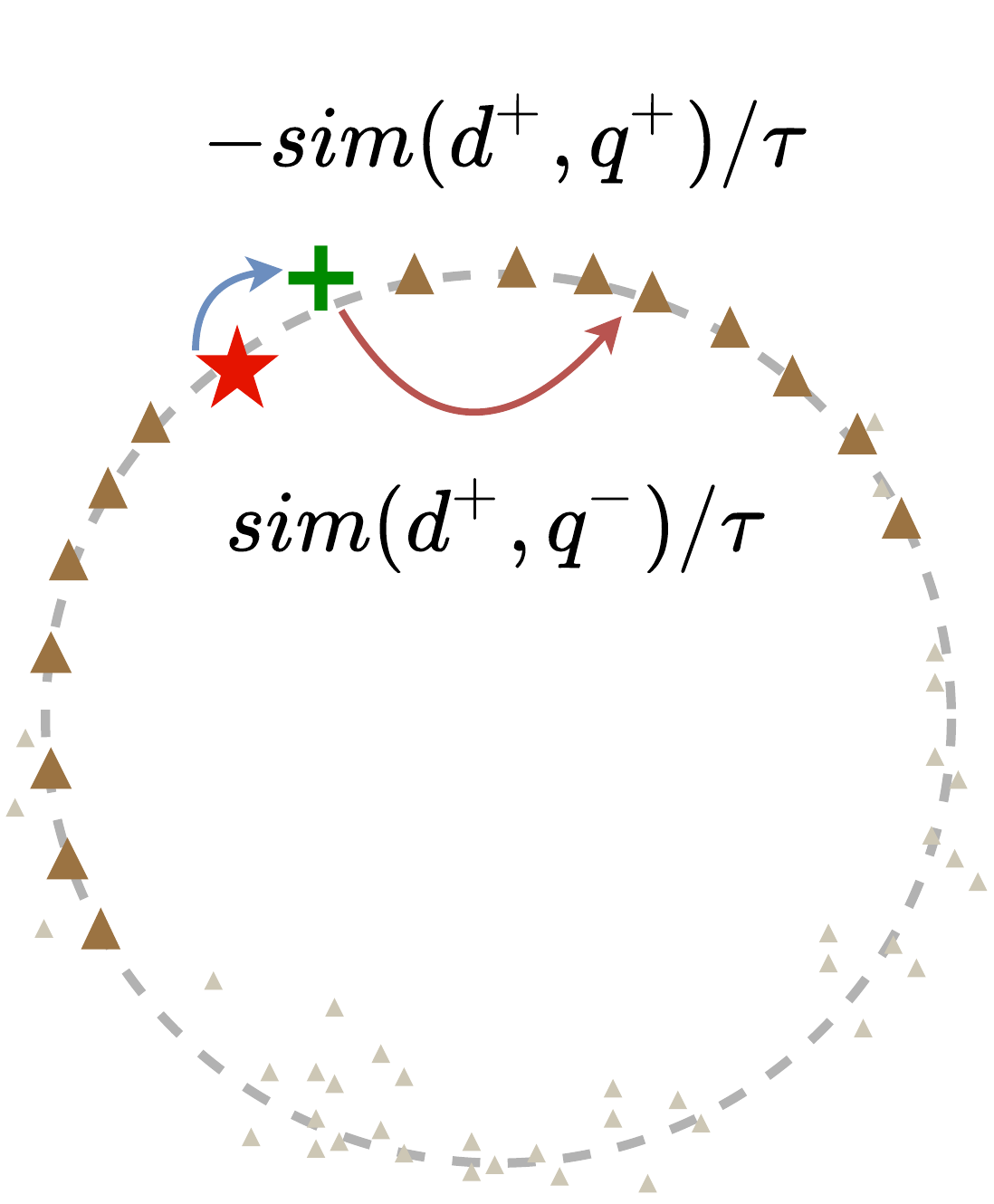}
    } %\hfill    
    \subcaptionbox{
        Contrastive Dual Learning. \label{fig:loss_insight_d}
    }{
        \includegraphics[width=0.225\textwidth]{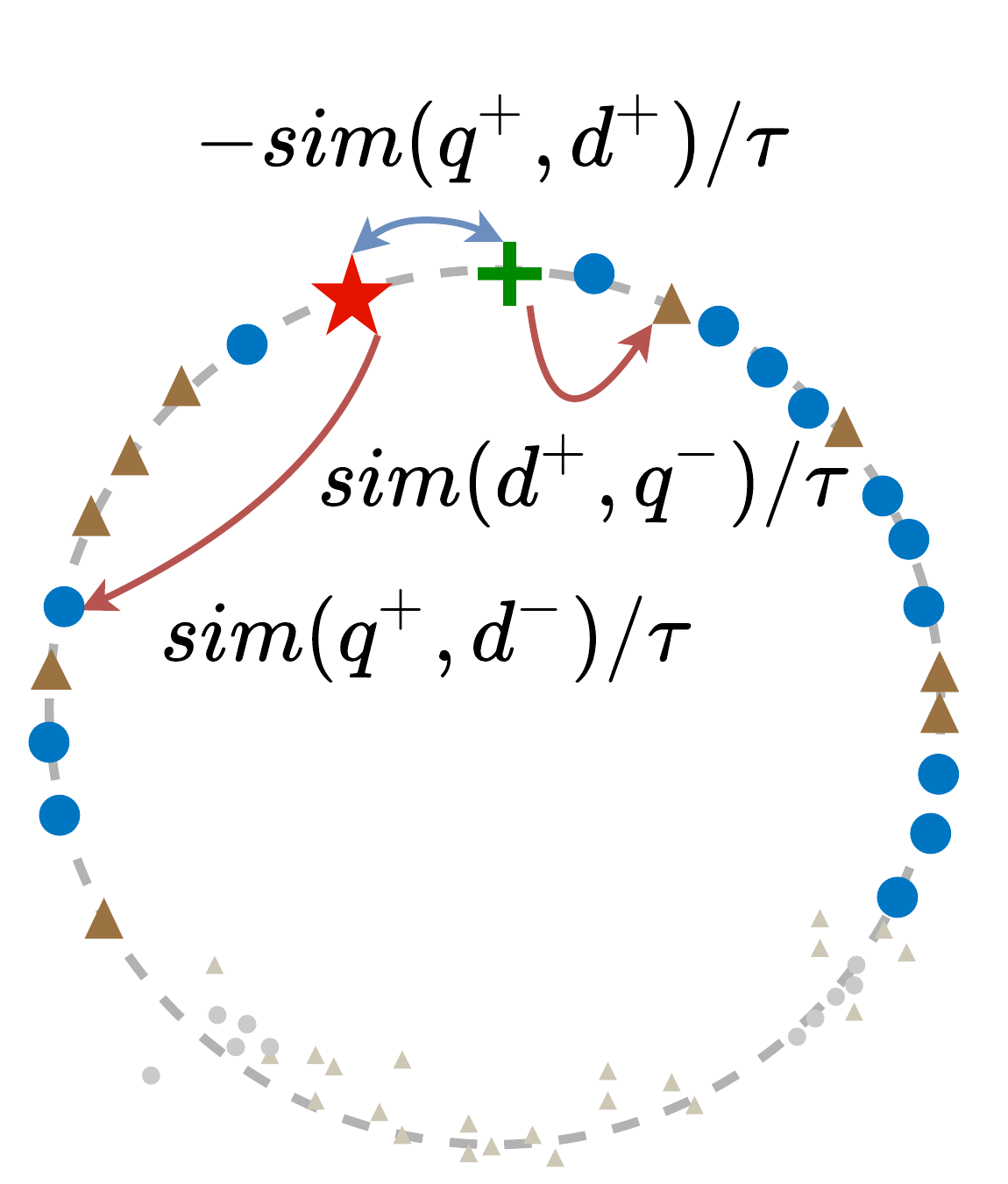}
    } %\hfill    
            
    \caption{The Illustration of Contrastive Dual Learning for Approximate Nearest Neighbor (\texttt{DANCE}) Training Paradigm.} 
    
    \label{fig:loss_insight}
\end{figure*}

In existing dense retrievers, the document embeddings are trained more sufficiently with the contrastive training of document retrieval task. Thus, some work fixes the learned document embeddings and focuses on optimizing the query representations. Such technology is leveraged to achieve better retrieval performance~\cite{zhan2020learning} or conduct an effective retriever in language modeling~\cite{guu2020realm} and question answering system~\cite{lewis2020retrieval}, demonstrating the importance of learning tailored query embeddings in dense retrievers.

To learn more fine-grained query representations, the query retrieval task provides an opportunity, which is inspired by the IR training axiom query likelihood~\cite{lavrenko2017lm}. Contrastively training dense retriever on the query retrieval task further optimizes the query embedding space more smooth and uniform, which is a dual task for the document retrieval task and can be incorporated in the model optimization with dual learning~\cite{he2016dual,xia2017dual,yi2017dualgan,lin2018conditional,xia2018model}. Dual learning is a method to train models of symmetric structures and has shown promising performance in some NLP tasks, such as machine translation and image caption~\cite{he2016dual,yi2017dualgan, tang2017question}. Xia et al.~\cite{xia2018model} further categorize dual learning algorithms as task-level dual learning and model-level dual learning according to whether the prime and dual model share components. They illustrate that model-level dual learning performs better compared to task-level dual learning~\cite{xia2018model}. Query retrieval and document retrieval are in the symmetric training directions and recent dense retrievers~\cite{xiong2020approximate} usually share the same BERT encoder for queries and documents. Thus, for dense retrievers, training with query retrieval and document retrieval can be regarded as a model-level dual learning paradigm.

\section{Methodology}
This section describes our proposed training paradigm for dense retrieval, Contrastive Dual Learning for Approximate Nearest Neighbor (\texttt{DANCE}), as shown in Figure~\ref{fig:loss_insight}. We first introduce the preliminaries of dense retrieval (Sec.~\ref{model:preliminary}), and then theoretically analyze how the contrastive training optimizes the query and document representations (Sec.~\ref{model:contrastive}). Finally, we describe our contrastive dual learning mechanism in \texttt{DANCE} (Sec.~\ref{model:dual}).

\subsection{Preliminary}\label{model:preliminary}
% 1. task definition and framework
% 2. negative sample learning.
% 3. dynamically updated negatives.
Given a query $q$ and a document collection $D=\{d_1,\dots,d_j,\dots,d_n\}$, dense retrievers calculate the ranking score $f(q,d)$ by learning the representations for documents similar to the intuition of document likelihood $p(d|q)$.

Dense retrievers~\cite{xiong2020dense,xiong2020approximate,karpukhin2020dense} leverage the same BERT encoder to get the representations of queries and documents:
\begin{align}
    H(q) &= \text{BERT}(\text{[CLS]} \circ  q \circ \text{[SEP]}); \\
    H(d) &= \text{BERT}(\text{[CLS]} \circ d \circ \text{[SEP]}),
\end{align}
where $\circ$ is the concatenation operation. ``[CLS]'' and ``[SEP]'' are special tokens in BERT. The representations of the first token ``[CLS]'' are used as representations of query $q$ and document $d$, which are denoted as $H_0(q)$ and $H_0(d)$, respectively. Then the similarity score $f(q,d)$ of query $q$ and document $d$ can be calculated with their dense representations:
\begin{equation}
  f(q,d)=sim(H_0(q),H_0(d)),
\end{equation}
where $sim(\cdot)$ is the similarity function to estimate the relevance between two embeddings. The efficient calculation of $sim(\cdot)$ for large scale dataset in our method is provided by FAISS\footnote{\url{https://github.com/facebookresearch/faiss}} and the dot product is usually used as the similarity function~\cite{karpukhin2020dense,xiong2020dense,xiong2020approximate}.

Then we can calculate the ranking loss $L$ with both positive document $d^+$ and negative document $d^-$ for the given query $q$ to contrastively train BERT:
\begin{equation}
    L= \sum\limits_{q}{
        \sum\limits_{d^+\in D^+}{
           l(q,d^+,D^-)
        }
    },
\end{equation}
where $D^+$ is the positive document collection for the given query $q$ that provided by annotation data.
$l(q,d^+,D^-)$ is the contrastive training loss function, which is the same as the state-of-the-art dense retrievers~\cite{xiong2020approximate}:
\begin{equation}
    l(q,d^+,D^-) = -\log\frac{e^{f(q^+,d^+)}}
    {e^{f(q^+,d^+)} + \sum\limits_{d^-\in D^-}{e^{f(q^+,d^-)}}},\label{eq:crossentropy}
\end{equation}
where $D^-$ is the collection of negative documents for query $q$ sampled with different retrieval methods, such as BM25~\cite{karpukhin2020dense} and dense retriever itself~\cite{xiong2020approximate}.

% The rest of this section describes how contrastive learning optimizes the embedding space of queries and documents (Sec.~\ref{model:contrastive}), then intuitively introduces how the proposed contrastive dual learning optimize the embedding space for retrieval (Sec.~\ref{model:dual}).

\subsection{Alignment and Uniformity}\label{model:contrastive}
% normalized temperature-scaled cross-entropy
% 1. contrastive learning insight
% 2. length information of the vectors -> angle information focus
Dense retrievers are contrastively trained to encode queries and documents with sampled negative documents as well as keep beneficial properties of the embedding space. As shown in recent research~\cite{wang2020understanding}, contrastive loss keeps ``alignment'' and ``uniformity'' of the document embedding space for retrieval. Such optimization progress is visualized in Figure~\ref{fig:loss_insight_a} and Figure~\ref{fig:loss_insight_b}.

Following previous work~\cite{chen2020simple}, we normalize the embeddings with L2 norm to keep representations of queries and documents in the unit hyperspherical space and calculate the similarity between the query $q$ and document $d$:
\begin{equation}
    f_{\text{norm}}(q,d)=sim( \left\|H_0(q)\right\|_2, \left\|H_0(d)\right\|_2),
\end{equation}
where $ -1\leq f_{\text{norm}}(q,d) \leq 1$ and $\left\|\cdot \right\|_2$ is the L2 normalization operation. The similarity function $f_{\text{norm}}(q,d)$ only focuses on estimating the relevance between query $q$ and document $d$ according to the angle of their representations. Then we use the normalized temperature-scaled cross entropy~\cite{chen2020simple} to replace the regular cross entropy function in Eq.~\ref{eq:crossentropy}:
\begin{equation}
    L_{\text{norm}}=-\log \frac{e^{f_{\text{norm}}(q,d^+)/\tau}}
    {e^{f_{\text{norm}}(q,d^+)/\tau} + \sum\limits_{d^-\in D_{\text{ANN}}^-}{ e^{f_{\text{norm}}(q,d^-)/\tau}}},\label{eq:scale_cross_entropy}
\end{equation}
where $D_{\text{ANN}}^-$ is the negative document collection. These documents are sampled from the ones distributed near to the query $q$ in the embedding space, which is the same as previous work~\cite{xiong2020approximate}. $\tau$ is the temperature hyperparameter used to control the sharpness of the softmax distribution.

Following Wang et al.~\cite{wang2020understanding}, the normalized temperature-scaled loss can be transformed as:
\begin{equation}
\begin{aligned}
    L_{\text{norm}}= 
    &\underbrace{- f_{\text{norm}}(q,d^+)/\tau}_{\text{alignment}} + \\ 
    &\underbrace{\log(e^{f_{\text{norm}}(q,d^+)/\tau} + \sum\limits_{d^-\in D_{\text{ANN}}^-}{e^{f_{\text{norm}}(q,d^-)/\tau})}}_{\text{uniformity}},\label{eq:align_uniform}
\end{aligned}
\end{equation}
where the loss function encourages the model to optimize embedding distributions and keeps two important properties of the embedding space:
\begin{itemize}
    \item \textbf{Alignment}: the query $q$ is distributed closer to the document $d$ than the negative document $d^-$ in the embedding space;
    \item \textbf{Uniformity}: the embedding distribution is encouraged to be uniform on the hyperspherical space by pushing the negative documents $d^-$ away from the given query $q$.
\end{itemize}
During contrastive training, dense retriever keeps alignment and uniformity of the embedding space, which helps to learn discriminative representations for documents.

\subsection{Contrastive Dual Learning}\label{model:dual}
% dual learning with doc&query likelihood
To further learn a uniform and smooth embedding space for queries, we train dense retrievers with a model-level dual learning task~\cite{xia2018model}, which optimizes dense retrievers with query retrieval task and document retrieval task. 
% They require the model to retrieve related queries or documents with given documents or queries, which are inspired by query likelihood $P(q|d)$ and document likelihood $P(d|q)$.
They are inspired by query likelihood $P(q|d)$ and document likelihood $P(d|q)$.

Query likelihood and document likelihood learning are symmetric. And the query likelihood $P(q|d)$ can also be approximated to the relevance between query and document:
\begin{equation}
    P(d|q) = P(q|d)\cdot P(d),
\end{equation}
where the document distribution $P(d)$ can be regard uniformly. 
To learn the query likelihood with contrastive training, we use a dual loss function $L_{\text{dual}}$:
\begin{equation}
    L_{\text{dual}}=-\log\frac{e^{f_{\text{norm}}(d,q^+)/\tau}}
    {e^{f_{\text{norm}}(d,q^+)/\tau} + \sum\limits_{q^-\in Q_{\text{ANN}}^-}{e^{f_{\text{norm}}(d,q^-)/\tau}}},
    \label{eq:dual_loss}
\end{equation}
where $Q_{ANN}^-$ is the negative query collection and sampled from the documents distributed near to the document $d$ in the embedding space. Similar to Eq.~\ref{eq:align_uniform}, the contrastive learning for query likelihood also keeps alignment and uniformity for the query embedding space:
\begin{equation}
\begin{aligned}
    L_{\text{dual}}= 
    &\underbrace{- f_{\text{norm}}(d,q^+)/\tau}_{\text{alignment}} + \\ 
    &\underbrace{\log(e^{f_{\text{norm}}(d,q^+)/\tau} + \sum\limits_{q^-\in Q_{\text{ANN}}^-}{e^{f_{\text{norm}}(d,q^-)/\tau})}}_{\text{uniformity}},
    \label{eq:dual_align_uniform}
\end{aligned}
\end{equation}
% &\underbrace{\log(e^{f_{\text{norm}}(d,q^+)/\tau} + \sum{e^{f_{\text{norm}}(d,q^-)/\tau})}}_{\text{uniformity}},
Finally, we add the prime training loss $L_{\text{norm}}$ and the dual training loss $L_{\text{dual}}$ to conduct our contrastive dual learning loss and train dense retrievers:
\begin{equation}
    L_{\text{final}}= L_{\text{norm}} + \lambda L_{\text{dual}},
\end{equation}
where $\lambda$ is used to weight the training loss of the dual task. 

% \subsection{Optimization Derivation}
% \input{formulas/ance_deriviate}
% \input{formulas/norm_deriviate}

% \input{tables/dataset_marco}
\section{Experimental Methodology} \label{sec:experiment_method}
This section describes the dataset, evaluation metrics, baselines, and experimental details of our implementation.

% \textbf{Dataset.} In all experiments, we use MS MARCO~\cite{bajaj2016ms} to evaluate model performance. The dataset consists of massive anonymized questions sampled from Bing’s search query logs and 8,841,823 passages extracted from 3,563,535 Bing retrieved web documents. 
\textbf{Dataset.} In all experiments, we use MS MARCO~\cite{bajaj2016ms} to evaluate model performance. The dataset consists of massive anonymized questions sampled from Bing’s search query logs and texts extracted from 3,563,535 Bing retrieved web pages.
For each query, the documents that have at least one related passage are recognized as relevant ones. We focus on the full ranking setting of the document retrieval task and retrieve 100 documents from the whole collection of 3,213,835 documents for each query to directly evaluate the retrieval performance. We keep the same official data partitions, the training set, development set and evaluation set contain 367,013 queries, 5,193 queries and 5,793 queries, respectively.

% \textbf{Evaluation Metrics.} 
\textbf{Evaluation Metrics.} Following the official evaluation of MS MARCO~\cite{bajaj2016ms}, we use evaluation metrics NDCG@10 and MRR@100 in our experiments, where MRR@100 is regarded as our main evaluation. The evaluation of MS MARCO document ranking is a black-box testing. All the results of submitted runs are available on the leaderboard\footnote{\url{https://microsoft.github.io/msmarco/}}. Statistic significance is tested by permutation test with $p <0.05$.

\textbf{Baselines.} 
%document ranking leaderboard, \texttt{Anserini's BM25}~\cite{yang2018anserini, nogueira2019doc2query}, \texttt{DE-HYBRID}~\cite{luan2020sparse}, \texttt{ME-HYBRID}~\cite{luan2020sparse}, and \texttt{ANCE}~\cite{xiong2020approximate}. Besides, to better evaluate the performance of \texttt{DANCE}, development set results from other recently proposed retrievers---\texttt{DPR} ~\cite{karpukhin2020dense, luan2020sparse}, \texttt{HDCT}~\cite{dai2020context}, \texttt{DE-BERT}~\cite{luan2020sparse} and \texttt{ME-BERT}~\cite{luan2020sparse}---are also listed.
% Our main baselines for retrieval-only setting include three published retrieval models that ranked top on the MS MARCO document ranking leaderboard, \texttt{Anserini's BM25}~\cite{yang2018anserini, nogueira2019doc2query}, \texttt{DPR} ~\cite{karpukhin2020dense, luan2020sparse} and \texttt{ANCE}~\cite{xiong2020approximate}.
% For retrieval-and-reranking setting, \texttt{ColBERT}~\cite{khattab2020colbert}, \texttt{PROP}~\cite{ma2020prop}, \texttt{LCE w. HDCT}~\cite{gao2021rethink} and \texttt{ANCE}~\cite{xiong2020approximate} are used to compare, where all of the results are produced with a single BERT-like reranker model. In both settings, \texttt{ANCE (FirstP)} is our main baseline.
Our baselines include two kinds of models: sparse retrieval models and dense retrieval models. The state-of-the-art dense retriever \texttt{ANCE (FirstP)} is regarded as the main baseline in our experiments.

The sparse retrieval baselines include \texttt{docT5query}~\cite{yang2018anserini, nogueira2019doc2T5query} and \texttt{HDCT}~\cite{dai2020context}, both of which inherit the discrete bag-of-word matches of the classical information retrieval method BM25 and focus on improving the document representations to achieve better retrieval performance for the sparse retriever. Different from vanilla BM25, \texttt{docT5query}~\cite{nogueira2019doc2T5query} improves document representations in sparse retrieval indexing by expanding documents with generated queries. \texttt{HDCT} uses pretrained language model, BERT~\cite{devlin2019bert} to predict the term weights in passages and then improves document bag-of-words representations by combining the passage term weights.

Other baselines are dense retrievers. They encode both queries and documents as dense representations and form an embedding space to conduct effective document retrieval~\cite{karpukhin2020dense,luan2020sparse,xiong2020approximate} with ANN search tools like ScaNN\footnote{\url{https://github.com/google-research/google-research/tree/master/scann}}~\cite{guo2020accelerating} and FAISS~\cite{johnson2019billion}. \texttt{DPR} first leverages contrastive training methods to train BERT~\cite{devlin2019bert} as the encoder for queries and documents. It proposes several negative document sampling methods, such as in-batch negatives, BM25 negatives and random negatives. \texttt{DPR w. BM25-Rand Neg} is compared in our experiments, which achieves the best retrieval performance among variants of \texttt{DPR}. To better train dense retrievers, \texttt{ANCE}~\cite{xiong2020approximate} uses RoBERTa~\cite{liu2019roberta} to learn representations of queries and documents and then samples negatives from the documents located near to the queries in the embedding space. Such training method avoids the diminishing gradients during training and helps to achieve competitive retrieval performance. Besides, two dense retrieval models, \texttt{DE-BERT} and \texttt{ME-BERT}, from the previous work~\cite{luan2020sparse} are compared in our experiments. \texttt{DE-BERT} is similar to \texttt{DPR}, and it uses the ``[CLS]'' hidden states to represent queries and documents. Different from \texttt{DE-BERT}, \texttt{ME-BERT} represents documents with multiple embeddings of the tokens of different positions learned by pretrained language models. The two models \texttt{DE-HYBRID} and \texttt{ME-HYBRID} linearly incorporate retrieval scores from sparse retriever to \texttt{DE-BERT} and \texttt{ME-BERT} with learnable weights, which are also compared.

\textbf{Implementation Details.} In our experiments, we follow our main baseline \texttt{ANCE}~\cite{xiong2020approximate} and contrastively train \texttt{DANCE}. We globally sample negatives from the whole collection of queries and documents to train \texttt{DANCE} and asynchronously update ANN indexes with the latest saved checkpoint.

We conduct a two-stage training progress to train \texttt{DANCE} on the full-ranking document retrieval task of MS MARCO, including the embedding normalization stage and the dual training stage. 
In the embedding normalization stage, we start with the checkpoint of well-trained dense retriever \texttt{ANCE}\footnote{https://github.com/microsoft/ANCE/}, which uses RoBERTa~\cite{liu2019roberta} as encoder. Then we map the representations of queries and documents into a hyperspherical embedding space and normalize them with L2 normalization. To tune the sharpness of softmax distribution and better train dense retrievers, we replace the standard cross entropy loss with the normalized temperature-scaled loss and set the temperature $\tau$ as 0.01.
Then we leverage the dual training paradigm to train \texttt{DANCE}, where the additional dual task (query retrieval task) is incorporated and the dual training loss weight $\lambda$ is set to 0.1. 
The two models for ablation study, \texttt{ANCE w. Norm} and \texttt{ANCE w. Dual}, trained with individual steps are also evaluated in our experiments.

\begin{table}
  \caption{Overall Performance. All models are evaluated on the MS MARCO document retrieval task. Superscript $\dagger$ indicates statistically significant improvement over \texttt{ANCE (FirstP)}$^\dagger$.}
  \label{tab:overall}
%   \begin{tabular}{l| c c c}
% p{0.3\linewidth}
\resizebox{0.48\textwidth}{!}{
  \begin{tabular}{l|c c | c}
    \hline
     \multirow{2}{*}{\textbf{Model}} & \multicolumn{2}{c|}{\textbf{Dev}} & \textbf{Eval} \\
     \cline{2-4}
    % \hline
    & NDCG@10 & MRR@100 & MRR@100 \\
    \hline
    % \textbf{Retrieval-only} & \multicolumn{3}{c}{ } \\
    % \hline 
    docT5query \cite{yang2018anserini, nogueira2019doc2T5query} & - & 0.327 & 0.291 \\
    HDCT \cite{dai2020context}&  - & 0.300 & -  \\
    DE-BERT \cite{luan2020sparse}&  - & 0.288 & -  \\
    ME-BERT \cite{luan2020sparse}&  - & 0.330 & -  \\
    DE-HYBRID \cite{luan2020sparse}&  - & 0.313 & 0.287  \\
    ME-HYBRID \cite{luan2020sparse}&  - & 0.346 & 0.310  \\
    DPR w. BM25-Rand Neg \cite{karpukhin2020dense, luan2020sparse}  &  0.362 & 0.312 & -  \\
    ANCE (FirstP) \cite{xiong2020approximate} &  0.437 & 0.373 & 0.334  \\
    %ANCE (MaxP) \cite{xiong2020approximate} & 0.445 & 0.384  & 0.342  \\
    \hline
    DANCE (FirstP) & 0.447$^{\dagger }$ & 0.383$^{\dagger }$  & 0.341 \\
    % DANCE (MaxP) &-  & 0.460 & 0.397  & - \\
    % \hline
    % \textbf{Retrieval-and-Reranking} & \multicolumn{3}{c}{ } \\
    % \hline
    % PROP \cite{ma2020prop}  & - & 0.445  &0.397\\
    % BERT-m1 base \cite{boytsov2020flexible} &-  & 0.441 &0.396  \\
    % ColBERT (MaxP end2end) \cite{khattab2020colbert}  & - & 0.440  &0.384\\
    % LCE w. HDCT \cite{gao2021rethink} & - & 0.434  &0.382\\
    % ANCE (FirstP) w. BERT (FirstP) & - & 0.431  & 0.380 \\
    % %ANCE (MaxP) w. BERT (MaxP) & 0.498 & 0.432  &0.391 \\
    % % DANCE (FirstP) w. BERT(FirstP) (no coor) & - & 0.491 & 0.417  & - \\
    % DANCE (FirstP) w. BERT(FirstP)  & 0.494 & 0.428  & 0.385 \\ % dev report -> coor checkpoints seperate
    % DANCE (MaxP) w. BERT (MaxP) & -  & 0.512 & 0.444  & TODO\\

  \hline
\end{tabular}}
\end{table}

Our experiments mainly focus on the \texttt{FirstP} setting proposed in previous work~\cite{dai2019deeper} to evaluate retrieval effectiveness. In \texttt{FirstP} setting, queries and documents are truncated and padded to the sequences with the maximum lengths of 64 and 512, respectively. \texttt{DANCE} is optimized with LAMB~\cite{you2019large} optimizer with warming up step of 3000 and learning rate of 5e-6. The training batch sizes are set to 4 and 210 for training and inference, respectively. The gradient accumulation step is set to 2. For other experiments, we keep the same with \texttt{ANCE}~\cite{xiong2020approximate} and use the IndexFlatIP in the toolkit FAISS~\cite{johnson2019billion} to build the index for query and document embeddings during retrieval. Our final model is trained with 110k steps in the normalization step and 80k in the contrastive dual training. We train our models with 8 GeForce RTX 2080 Ti GPUs of 11GB with half-precision setting and the inference program runs on 4 same GPUs.

% For the BERT reranker, we employ the same setting in OpenMatch\footnote{\url{https://github.com/thunlp/OpenMatch}} and conduct BERT based reranking models following the retrieval results from \texttt{DANCE}. The negative training documents are 20 documents randomly sampled from the top-100 retrieval results of each training query with \texttt{DANCE}. During training, we use the AdamW optimizer, set the maximum query length as 64, maximum document length as 445, and pad the concatenation of query and document to 512. In reranker training, we use 4 GeForce RTX 2080 Ti GPUs, set the learning rate to $2e-5$, gradient accumulation steps to 4, number of warmup steps to 2000, and training batch size to 4. Besides, the features and scores from BERT reranker and retrieval scores of \texttt{DANCE} are ensembled with Coordinate-Ascent \cite{metzler2007linear} through 5-fold cross validation training on the development set.

\section{Evaluation Result}
In this section, we present six groups of experiments on the overall performance of \texttt{DANCE}, the embedding distributions of queries and documents learned by \texttt{DANCE}, the retrieval effectiveness in different testing scenarios and case studies.

\subsection{Overall Performance}
\begin{table}
  \caption{The Ranking Performance of Ablation Models of \texttt{DANCE} on MS MARCO Document Retrieval Task. Superscripts $\dagger, \ddagger, \mathsection $ indicate statistically significant improvements over \texttt{ANCE (FirstP)}$^\dagger$, \texttt{ANCE w. Norm (FirstP)}$^\ddagger$, and \texttt{ANCE w. Dual}$^ \mathsection$, respectively.}
  \label{tab:marco_doc}
  \begin{tabular}{ l| c c }
    \hline
    \multirow{2}{*}{\textbf{Model}} &  \multicolumn{2}{c}{\textbf{Dev}} \\
    % \cline{2-3}
    % & NDCG@10 & MRR@100 \\
    \cline{2-3}
     & NDCG@10 & MRR@100 \\
    \hline
    ANCE (FirstP) & 0.437 & 0.373 \\
    \hline
    ANCE w. Norm (FirstP) & 0.443$^{\dagger}$ & 0.380$^{\dagger}$ \\
    ANCE w. Dual (FirstP) & 0.444$^{\dagger}$ & 0.381$^{\dagger}$  \\
    DANCE (FirstP) & \textbf{0.447}$^{\dagger \ddagger}$ & \textbf{0.383}$^{\dagger \ddagger}$ \\
    % \hline
    % ANCE (MaxP) & 0.445 & 0.384 \\
    % DANCE w. Norm (MaxP) & 0.450 & 0.386 \\
    % DANCE (MaxP) & \textbf{0.460} & \textbf{0.397} \\
  \hline
\end{tabular}
\end{table}
% The performance of \texttt{DANCE} and baselines on MS MARCO document retrieval task is shown in Table~\ref{tab:overall}. The baselines are categorized into two groups: retrieval models and retrieval-reranking models.

% For document retrieval, \texttt{DANCE} improves \texttt{ANCE (FistP)} with 0.7\% MRR score, demonstrating the effectiveness of our contrastive dual training paradigm by carefully learning query representations. When we use the vanilla \texttt{BERT (FirstP)} model to rerank the candidate documents retrieved by \texttt{DANCE} and \texttt{ANCE}, \texttt{DANCE} shows consist improvement in the sole retrieval scenario comparing with \texttt{ANCE}, which illustrates the effectiveness of \texttt{DANCE} in the retrieval stage can also thrive on the overall document retrieval performance. Besides, \texttt{\texttt{DANCE} (FirstP) w. BERT(FirstP)} also outperforms two strong baselines \texttt{ColBERT (MaxP end2end)} and \texttt{LCE w. HDCT}. It further confirms the effectiveness of \texttt{DANCE} that only considers the first passage in the document for retrieval and reranking.

The performance of \texttt{DANCE (FirstP)} and baselines on MS MARCO document retrieval task is shown in Table~\ref{tab:overall}.

\texttt{DANCE (FirstP)} outperforms the sparse retrievers \texttt{docT5query} and \texttt{HDCT}, showing the effectiveness of well-trained dense retrieval models by conducting semantic matches in document retrieval.
Among all the dense retrievers, both \texttt{DANCE (FirstP)} and \texttt{ANCE (FirstP)} show much better performance by choosing more valuable negative documents to contrastively train dense retrievers. 
Benefited by the dual training paradigm, \texttt{DANCE (FirstP)} further improves \texttt{ANCE (FirstP)} with 0.7\% MRR@100 score. The improvement demonstrates that the additional query retrieval task can help dense retrievers learn more tailored representations of queries and documents.

\subsection{Ablation Study}
% \input{tables/marco_passage_dev_small}
% \input{tables/marco_dev_merge}
% compound > single
In this part, we conduct ablation studies to further explore the roles of different components in \texttt{DANCE} on MS MARCO document retrieval task, as shown in Table~\ref{tab:marco_doc}.

The two individual modules of \texttt{DANCE}, hyperspherical normalization (\texttt{Norm}) and contrastive dual learning (\texttt{Dual}), are evaluated in this experiment, which are two optimization strategies used in \texttt{DANCE} to train dense retrievers. \texttt{Norm} and \texttt{Dual} focus on optimizing the embedding space from different aspects. The \texttt{Norm} module normalizes the embedding space and maps query and document representations into a unit hyperspherical space with L2 normalization, making the similarity calculation mainly focusing on the angle between two vectors.
And the \texttt{Dual} module incorporates the additional training object, query retrieval task, in the training process and mainly optimizes query embedding space for text retrieval.

We first evaluate the retrieval performance of two individual modules of \texttt{DANCE}, \texttt{Norm} and \texttt{Dual}, on the document retrieval task. Compared with baseline \texttt{ANCE (FirstP)}, \texttt{ANCE w. Norm  (FirstP)} and \texttt{ANCE w. Dual (FirstP)} achieve about 0.7\% and 0.8\% improvements of MRR@100 score on the development set, demonstrating their effectiveness on learning a more tailored embedding space for text retrieval from different aspects.
With both individual modules incorporated into the dense retriever, the performance of \texttt{DANCE (FirstP)} is further improved and gets the best ranking performance among all models.

\begin{figure}[t]
  \centering
  \includegraphics[width=0.35\textwidth]{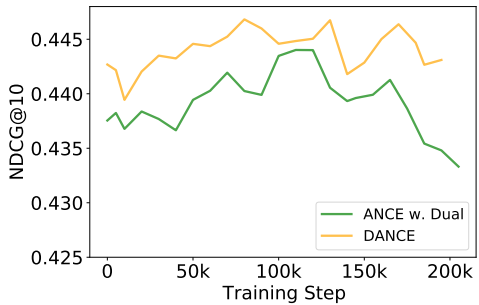}
  \caption{Document Retrieval Performance of \texttt{ANCE w. Dual} and \texttt{DANCE} during Training. The model performance is evaluated on the development set.}
  \label{fig:dance_overfit}
\end{figure}

In our experiments, we find that \texttt{Norm} can alleviate the overfitting problem during training. As shown in Figure~\ref{fig:dance_overfit},
\texttt{DANCE} shows a more stable performance compared with \texttt{ANCE w. Dual} in the training process. Some dense retrieval models also face the unstable training problem, which encourages them to use BM25 negatives to warm up training~\cite{xiong2020approximate} and carefully optimize the embedding space locally with lots of tricks~\cite{lewis2020pre}. The \texttt{Norm} module conducts more stable training progress and may shed some light to deal with these problems. To evaluate the effectiveness of contrastive dual training, we regard the \texttt{ANCE w. Norm  (FirstP)} model as our main baseline in the following experiments.

\subsection{Learned Embedding Space of \texttt{DANCE}}\label{sec:embedding_space}
% 0. better space -> better q2d retrieval
% 1. Document space alignment up +  Document space uniformity up
% 2. Query Space uniformity up. 
% 3. summary: additional proof -- Fig.3 pairwise distance numerically up. improve the uniformity of both space. according to Wang et. al, better unif -> better contrastive learning -> better retrieval

This set of experiments further explores the embedding distributions of queries and documents learned by \texttt{DANCE}.

\textbf{Pairwise Distance of Learned Embeddings.} As shown in Table~\ref{tab:mean_dist_dev}, we first evaluate the mean and variance of the cosine distances between query-query pairs, document-document pairs, and query-document pairs in the hyperspherical embedding spaces learned by \texttt{ANCE w. Norm} and \texttt{DANCE}. We use all queries from both training and development sets in this experiment.

\begin{table}
  \caption{The Statistics of Embedding Distance on MS MARCO Document Retrieval Task. The mean value and variance value of embedding distances between query-query pairs, document-document pairs and query-document pairs in the embedding space are calculated. The mean value represents the density of embedding distribution. The variance value represents the uniformity and smooth of the embedding space.}
  \label{tab:mean_dist_dev}
  \begin{tabular}{l| c c|c }
    \hline
    \multirow{2}{*}{\textbf{Distance Pair}} & \multicolumn{2}{c|}{\textbf{Mean Distance}} & \textbf{Variance} \\
    \cline{2-3}
    & ANCE w. Norm & DANCE & \textbf{Change} \\
    \hline
% outlier
    % Doc-Doc & 0.177 & 0.190  & - 3.11e-5 \\
    % Que-Que & 0.294 & 0.276  & - 2.05e-4 \\
    % Que-Doc & 0.250 & 0.243 &  - 3.20e-4 \\
    Doc-Doc & 0.179 & 0.191  & - 1.22e-5 \\
    Que-Que & 0.299 & 0.281  & - 5.00e-4 \\
    Que-Doc & 0.251 & 0.243 & - 3.71e-4  \\
  \hline
\end{tabular}
\end{table}

%  dev to all, below
% \begin{table}
%   \caption{Distance Comparison on MS MARCO Document Devset. We use the embeddings of 5,193 queries and their corresponding 5,185 positive documents in MS MARCO development to calculate the pairwise cosine distance.}
%   \label{tab:mean_dist_dev}
%   \begin{tabular}{l| c c|c }
%     \hline
%     \multirow{2}{*}{\textbf{Distance Pair}} & \multicolumn{2}{c|}{\textbf{Mean Distance}} & \textbf{Variance} \\
%     \cline{2-3}
%     & DANCE w. Norm & DANCE & \textbf{Change} \\
%     \hline
%     Doc-Doc & 0.180 & 0.193  & - 4.93e-5 \\
%     Que-Que & 0.294 & 0.275  & - 2.10e-4 \\
%     Que-Doc & 0.251 & 0.243 &  - 3.96e-4 \\
%   \hline
% \end{tabular}
% \end{table}
By evaluating the change of pairwise distances before and after adding the \texttt{Dual} module, we find that \texttt{DANCE} shows a statistical difference of embedding distributions of \texttt{ANCE w. Norm} and \texttt{DANCE}. 
First, compared with \texttt{ANCE w. Norm}, the average distance of document-document pairs becomes larger in the embedding space learned by \texttt{DANCE}. It shows the document embeddings learned by \texttt{DANCE} are more scattered, benefiting the document retrieval in the embedding space. Meanwhile, the mean distance of query-document pairs shows a contrary trend. \texttt{DANCE} reduces it and generally concentrates the query embeddings closer to the document embedding population, apparently leading to a smaller mean distance of query-query pairs. As shown in the next experiment, such document embedding distribution derives from the concentrated query embedding distribution and the contrastive training on document retrieval task.

For all three kinds of pairwise distances, the consistently reducing variance value in \texttt{DANCE} further manifests the learned embedding spaces of queries and documents are more uniform and smooth, which is one of the sources of the effectiveness of \texttt{DANCE}.

% \begin{figure*}[h]
%   \centering
%   \includegraphics[width=.24\textwidth]{pictures/dev-q-804_pos-qid-145935_ann-neg-D-50_random-neg-D-50_ppl20_ANCE.pdf}
%   \includegraphics[width=.24\textwidth]{pictures/dev-q-804_pos-qid-145935_ann-neg-D-50_random-neg-D-50_ppl20_normedANCE.pdf}
%   \includegraphics[width=.24\textwidth]{pictures/dev-q-804_pos-qid-145935_ann-neg-D-50_random-neg-D-50_ppl20_normedANCE-dual.pdf}
%   \includegraphics[width=.24\textwidth]{pictures/dev-q-3482_pos-d-312985_longtail_normANCE_dANCE__top10-neg-Q-dist_ppl20_lr100_plots_IP-distance.pdf}
%   \caption{q2d retrieval, q+ (red star), d+ (green cross), retrieval D- (blue dot), random D-(grey dot); ANCE, normalized ANCE, normalized ANCE + dual learning; t-SNE visualization with perplexity 20}
%   \Description{--}
% \end{figure*}

\begin{figure*}[t]
    \centering
    \setlength{\lineskip}{\medskipamount}
    
    \subcaptionbox{
        ANCE.\label{fig:emb_dist_a}
    }{
        \includegraphics[width=0.21\textwidth]{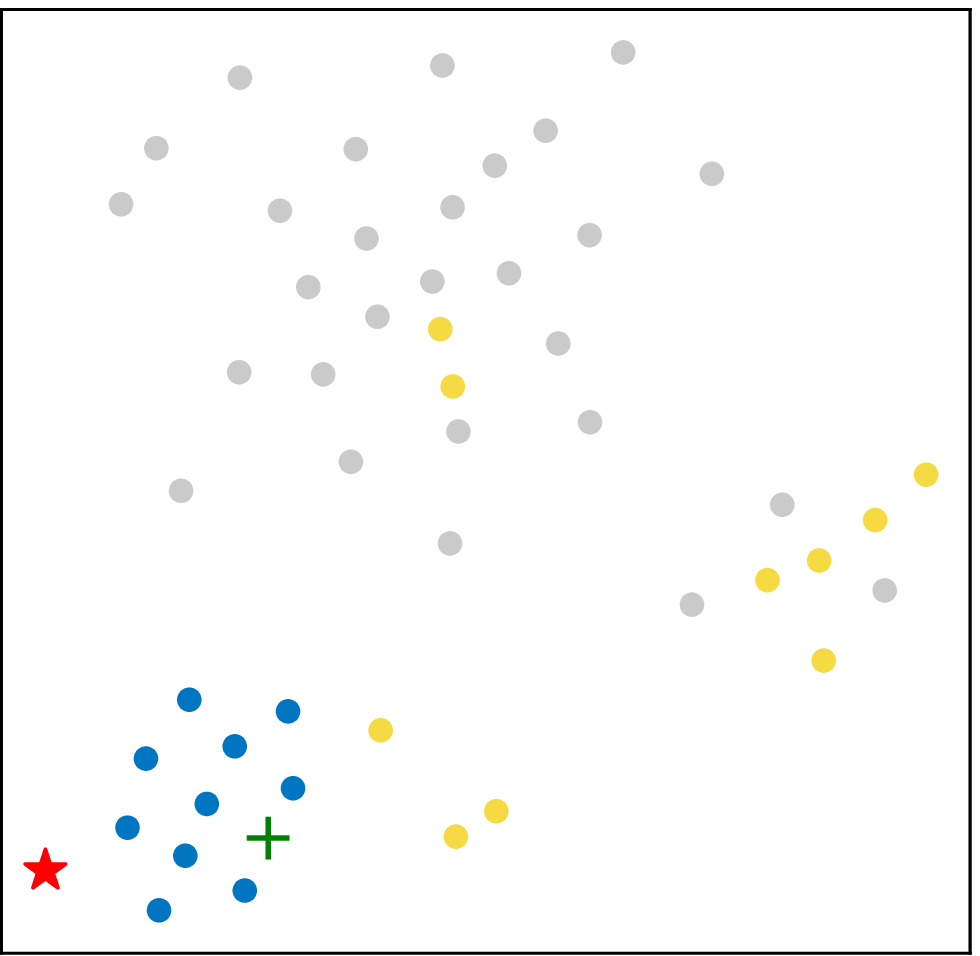}
    } %\hfill    
    \subcaptionbox{
        ANCE w. Norm.\label{fig:emb_dist_b} 
    }{
        \includegraphics[width=0.21\textwidth]{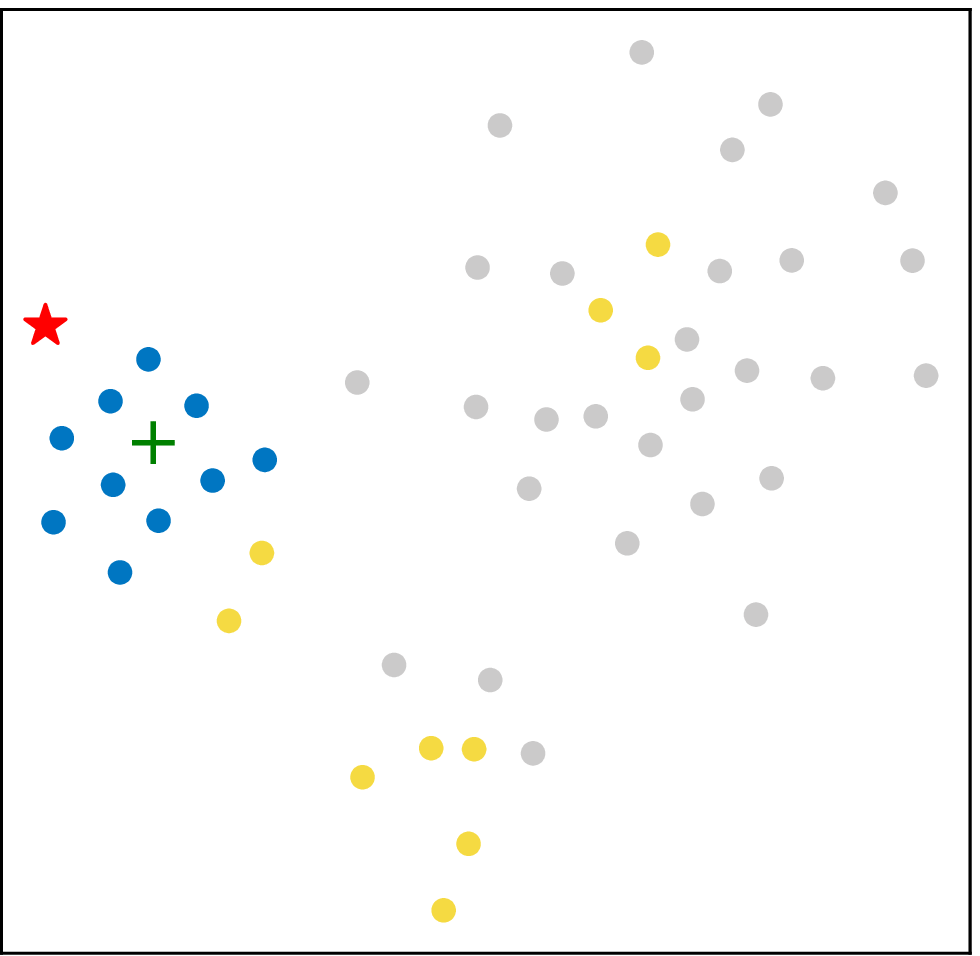}
    } %\hfill   
    \subcaptionbox{
        DANCE.\label{fig:emb_dist_c}
    }{
        \includegraphics[width=0.21\textwidth]{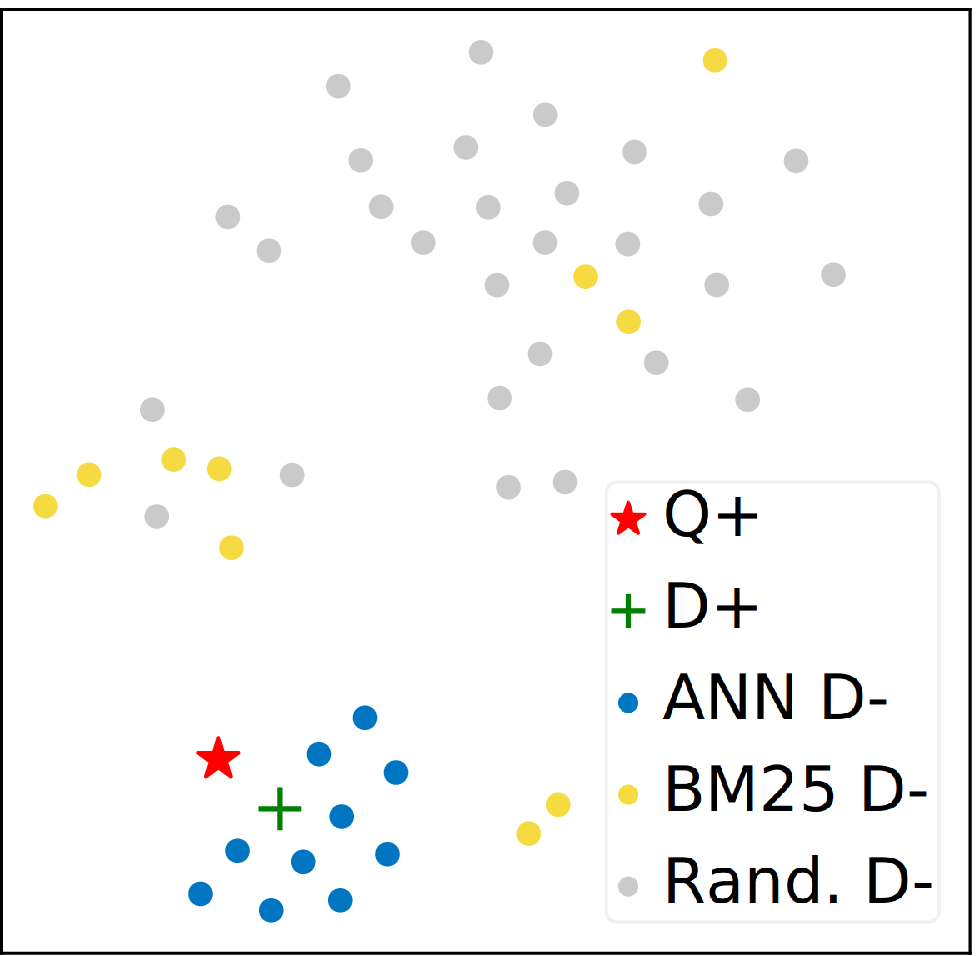}
    } %\hfill   
    \subcaptionbox{
        DANCE in Query Space.\label{fig:emb_dist_d}
    }{
     \includegraphics[width=0.21\textwidth]{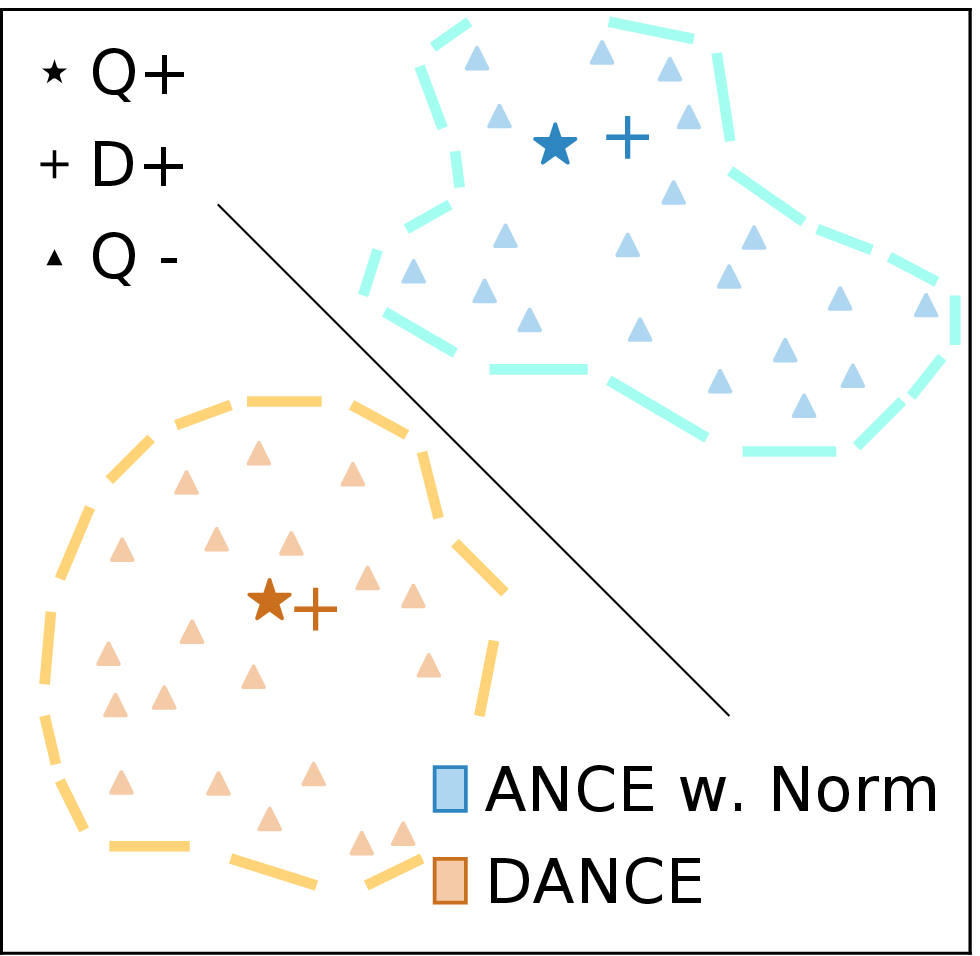} % ppl=10
        
    } %\hfill   
            
    % \caption{
    % Embedding Space Distributions of Different Dense Retrieval Models. In Figure~\ref{fig:emb_dist_a}, Figure~\ref{fig:emb_dist_b} and Figure~\ref{fig:emb_dist_c}, we choose the same case from the development set to visualize the embedding distribution of \texttt{ANCE}, \texttt{ANCE w. Norm} and \texttt{DANCE}, respectively. The documents consist of top-10 negative documents retrieved by individual model, top-10 negative documents retrieved by BM25 model and 30 documents that are randomly sampled from the whole document collection. In Figure~\ref{fig:emb_dist_d}, given a document, top-20 negative queries retrieved by \texttt{ANCE w. Norm} and \texttt{DANCE} and the query that is related to the given document are plotted to visualize the query embedding space.
    % } 
    \caption{
    Embedding Space Distributions of Different Dense Retrieval Models. In Figure~\ref{fig:emb_dist_a}, Figure~\ref{fig:emb_dist_b} and Figure~\ref{fig:emb_dist_c}, we choose the same case from the development set to visualize the embedding distribution of \texttt{ANCE}, \texttt{ANCE w. Norm} and \texttt{DANCE}, respectively. For the three figures on the left, visualized embeddings consist of a related query-document pair, top-10 negative documents retrieved by a dense retriever, top-10 negative documents retrieved by BM25 model and 30 documents that are randomly sampled from the whole document collection. In Figure~\ref{fig:emb_dist_d}, to visualize the query embedding space, a related query-document pair, top-20 negative queries retrieved by \texttt{ANCE w. Norm} and \texttt{DANCE} are plotted.
    } 
    % we use t-SNE visualization with perplexity 20.
    \label{fig:emb_dist}
\end{figure*}
\begin{figure}[t]
    \centering
    \setlength{\lineskip}{\medskipamount}
    \subcaptionbox{
        Document Retrieval.\label{fig:longtail_recalls_a}
    }{
        \includegraphics[width=0.2\textwidth]{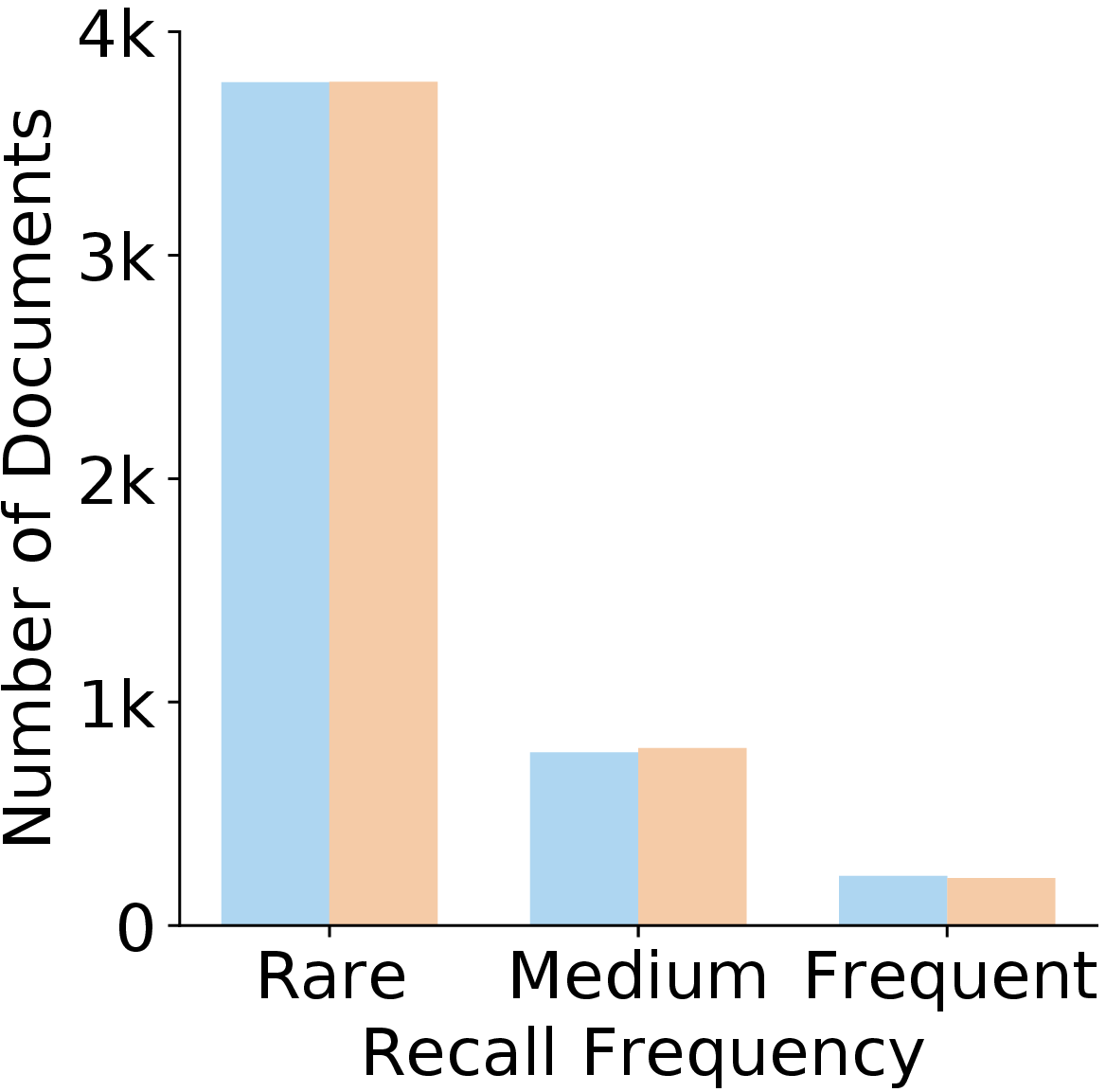}
    } %\hfill    
    \subcaptionbox{
        Query Retrieval.\label{fig:longtail_recalls_b}
    }{
        \includegraphics[width=0.2\textwidth]{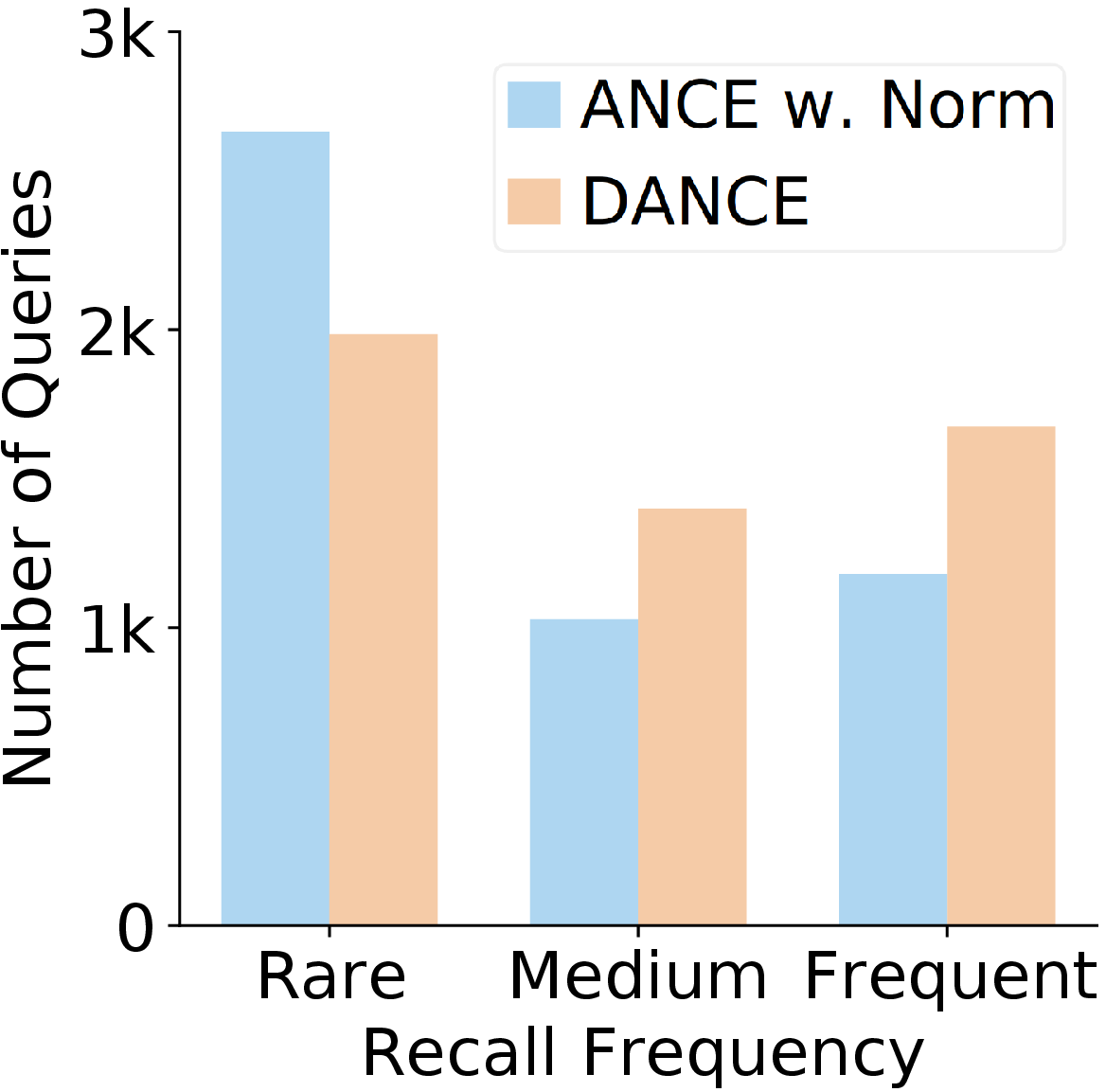}
    } %\hfill
    
    \caption{Number Distributions of Queries and Documents with Different Recall Frequency. All documents and queries in the development set are divided into three groups according to the recall frequency, which include rare, medium and frequent.} 
    \label{fig:longtail_recalls}
\end{figure}
% \vspace{0.5cm}
\begin{figure}[t]
    \centering
    \subcaptionbox{
        Document Retrieval.\label{fig:dist_recall_a}
    }{
        \includegraphics[width=0.21\textwidth]{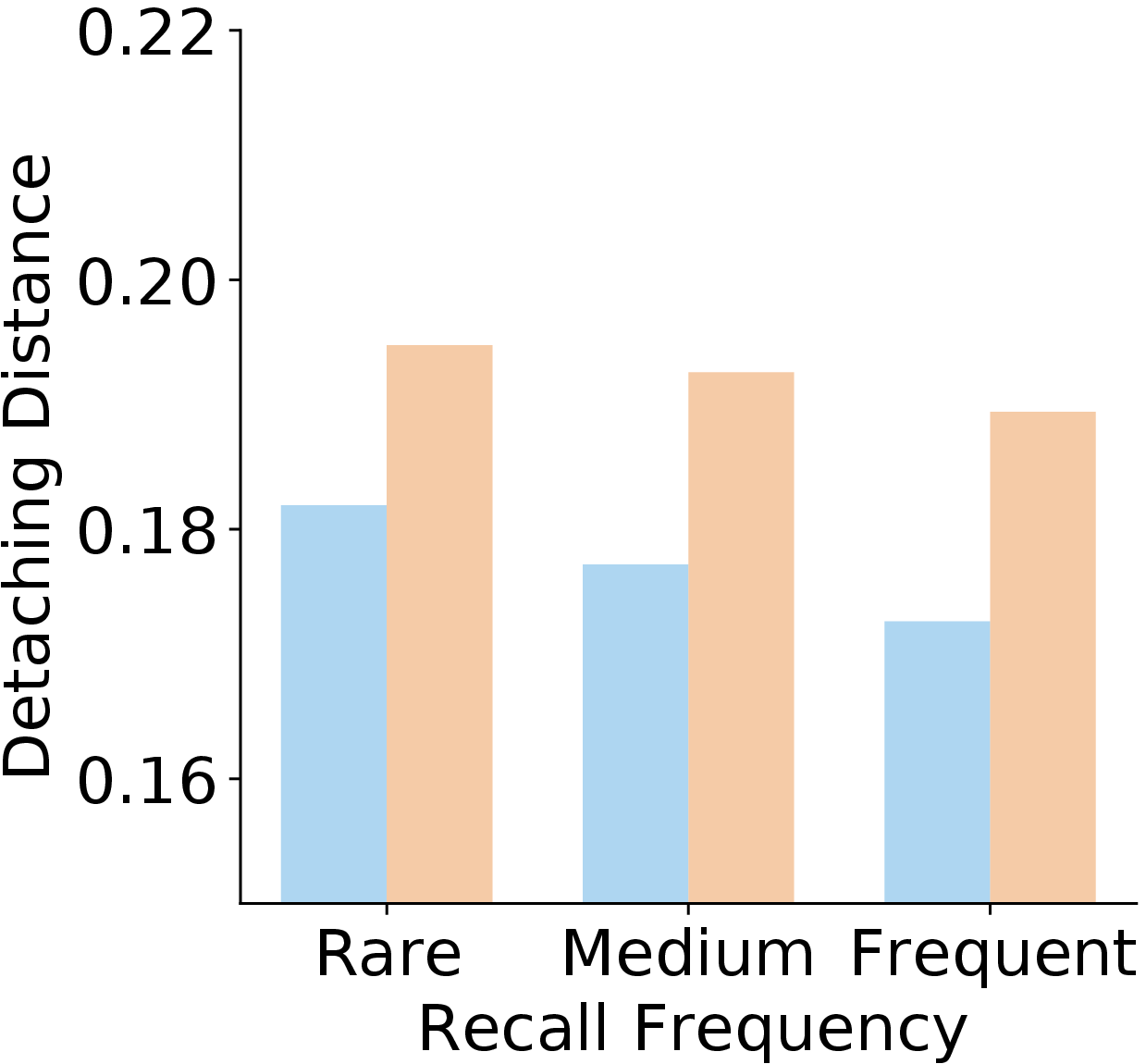}
    } \hfill    
    \subcaptionbox{
        Query Retrieval.\label{fig:dist_recall_b}
    }{
        \includegraphics[width=0.21\textwidth]{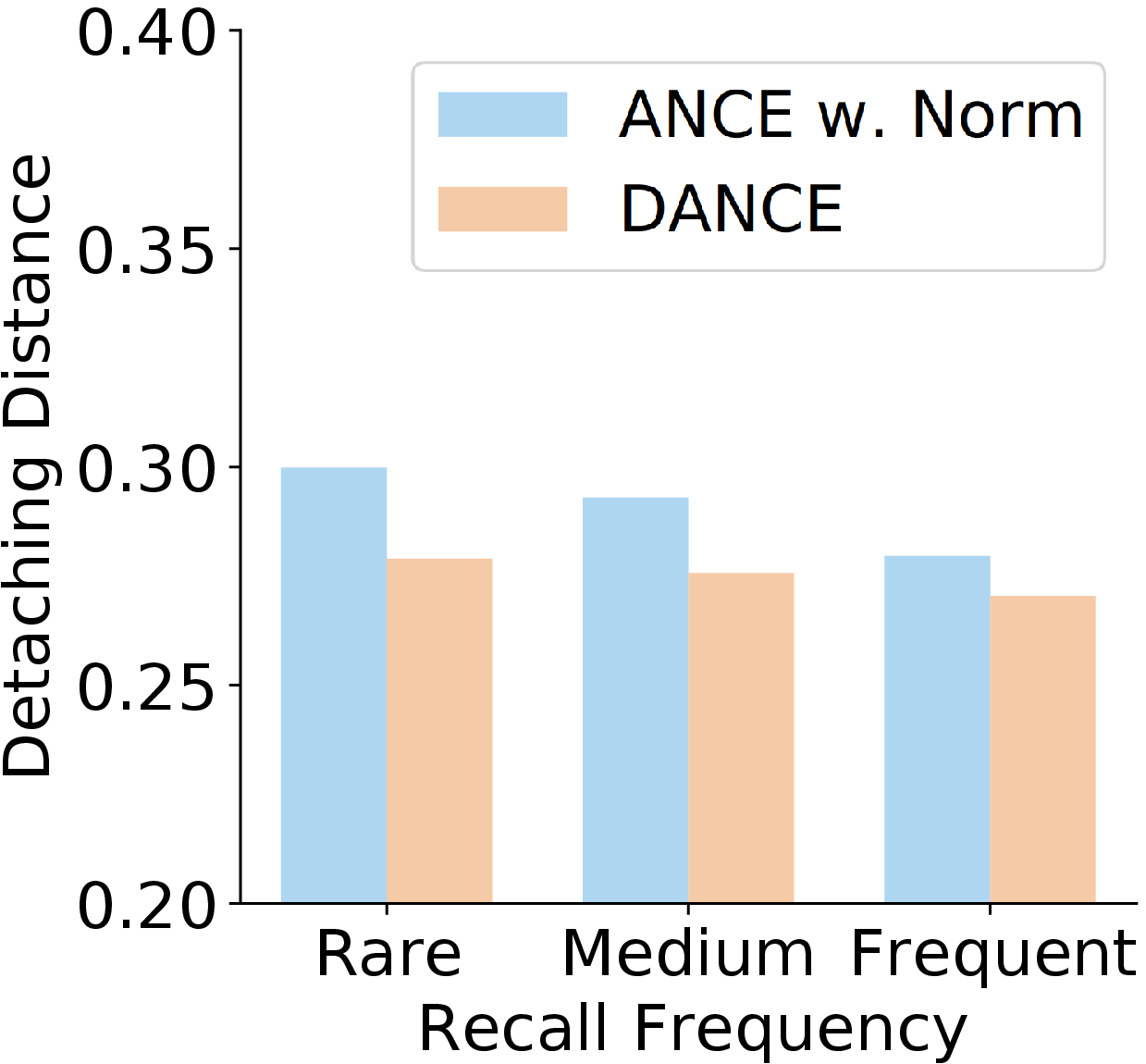}
    } \hfill
    
    \caption{Detaching Distances of Queries and Documents with Different Recall Frequency.
    The recall times of queries and documents are calculated by query retrieval task and document retrieval task. All queries and documents are divided into three groups according to the recall frequency, which include rare, medium and frequent.}
    \label{fig:dist_recall}
\end{figure}
% \vspace{0.5cm}
\textbf{Document Embedding Visualization.} In this experiment, we visualize the document embedding space via t-SNE in Figure~\ref{fig:emb_dist}.

We choose one matched query-document pair from the development set of MS MARCO document retrieval dataset and select some documents to plot the embedding space. These documents are chosen from top-ranked documents of the corresponding model, top-retrieved documents from BM25, and random sampling. The embedding spaces of three models, \texttt{ANCE}, \texttt{ANCE w. Norm} and \texttt{DANCE}, are shown in Figure~\ref{fig:emb_dist_a}, \ref{fig:emb_dist_b}, and \ref{fig:emb_dist_c}, respectively.

\texttt{DANCE} shows its ability to position queries and documents appropriately in the embedding space when training with the contrastive dual training paradigm. First, compared with \texttt{ANCE w. Norm}, \texttt{DANCE} can better align the query-document pairs by pulling the query embedding closer to the related document. Meanwhile, the surrounding unrelated documents are pushed away during contrastively training on the document retrieval task, making them more scattered and discriminative. 
Such document embedding distribution intuitively keeps the ``uniformity'' of embedding space and helps to distinguish confusable documents during retrieval.

Besides, different with vanilla \texttt{ANCE}, \texttt{ANCE w. Norm} only normalizes the representations of queries and documents and restricts them in a hyperspherical space.
As expected, \texttt{ANCE w. Norm} concentrates the embedding space and shares almost the same embedding distribution of \texttt{ANCE}.

\textbf{Query Embedding Distribution.} In the rest of this set of experiments, we select the same relevant query-document pair as previous experiments and retrieve top-ranked negative queries with corresponding models to visualize the query embedding distributions of \texttt{ANCE w. Norm} and \texttt{DANCE}. 

As shown in Figure~\ref{fig:emb_dist_d}, \texttt{DANCE} forms a more smooth and uniform query embedding space by better keeping the ``alignment'' and ``uniformity''.
First, same as our previous observation, \texttt{DANCE} positions the related query closer to the document in the learned embedding space, showing better ``alignment'' between matched query-document pairs. Second, \texttt{DANCE} learns a more homogeneous query embedding distribution derives from contrastive training on the additional query retrieval task, which helps to better maintain the ``uniformity'' of the query embedding space.

% Compared with \texttt{ANCE} visualized in Figure~\ref{fig:emb_dist_a}, embeddings in Figure~\ref{fig:emb_dist_b} shows that \texttt{ANCE w. Norm} only normalizes embeddings of queries and documents, which restricts these embeddings in a hyperspherical embedding space. 
% In Figure~\ref{fig:emb_dist_c}, compared with \texttt{ANCE w. Norm}, \texttt{DANCE} assigns locations for queries that are closer to document embeddings, which apparently reduces the mean distance values of query-document pairs and indirectly decreases the average query-query distances. 
% Thrived from the query embedding calibration supported by training with query retrieval task, the query is assigned a more appropriate position in the embedding space, which is closer to the golden document and potentially related documents retrieved by BM25, demonstrating the effectiveness of \texttt{DANCE} on learning query representations. Because the query embeddings are pulled closer to document embeddings, the main contrastive training loss of \texttt{DANCE}---document retrieval---stimulates dense retrievers to push surrounding unrelated documents away from the query and keeps the ``uniformity'' of document embedding space, which makes the average distance of document-document pairs larger in \texttt{DANCE} than the average distance of \texttt{ANCE w. Norm}.

\subsection{Embedding with Different Recall Frequency}\label{sec:results:recall}
In this subsection, we further explore how the proposed contrastive dual training paradigm optimizes the embedding distribution. We conduct two experiments to study the recall frequencies of queries and documents during contrastive training and the change of the embedding distributions with different recall frequencies.

% We use the recall frequency to estimate the chances of queries and documents to get optimized in dense retriever training. The probabilities of queries and documents that are sampled as negative ones in the contrastive training will be higher if they are more likely to be recalled. 
To estimate the probabilities of queries and documents that are sampled as negative ones in contrastive training, we use the recall frequency for approximation.
It calculates the times of queries and documents that appear in the top-100 retrieved candidates in the corresponding retrieval tasks, query retrieval, and document retrieval. All queries and documents in the development set are used in our experiments and divided into three groups according to the recall frequency, which includes rare (recalled once), medium (recalled twice), and frequent (recalled more than twice).

In the first experiment, we plot the number distributions of documents and queries along with different recall frequencies in Figure~\ref{fig:longtail_recalls}. The main difference between \texttt{ANCE w. Norm} and \texttt{DANCE} is on the query recall frequency distribution. As shown in Figure~\ref{fig:longtail_recalls_b}, \texttt{DANCE} assigns more uniform recall frequencies to queries, which balances the sampling probabilities of queries during contrastive training on the query retrieval task. Such sampling mechanism optimizes query representations more sufficiently, especially for those long-tailed queries. Next, we further study how the recall frequency influences embedding distributions.

% We study the embedding distributions of queries and documents with different recall frequency. 
In the second experiment, we first introduce the detaching distance to serve for embedding distribution analysis. The detaching distances of queries and documents are calculated by the mean distances between one query or document and others in the group of query or document.
The queries or documents with larger detaching distances indicate that they are located in a more scattered area and far from the query or document cluster.

Then we plot the detaching distances of queries and documents with different recall frequencies in Figure~\ref{fig:dist_recall}. 
Overall, the queries and documents with higher recall frequency usually have smaller detaching distances. 
The main reason is that the semantic meanings behind these representations are more confusable, leading to similar representations, concentrated embedding distributions and higher probabilities to be recalled in the contrastive training.
Compared with \texttt{ANCE w. Norm}, \texttt{DANCE} achieves larger detaching distances of document pairs and smaller detaching distances of query pairs. It again demonstrates that \texttt{DANCE} learns more scattered and discriminative document embeddings and concentrates the query embeddings in the embedding space, which are observed in previous experiment (Sec.~\ref{sec:embedding_space}).
Then we also conduct following experiments to study how the embedding distribution learned by \texttt{DANCE} affects the ranking effectiveness.

\begin{figure}[t]
    \centering
    \subcaptionbox{
        Query Retrieval.\label{fig:longtail_ndcg_a}
    }{
        \includegraphics[width=0.22\textwidth]{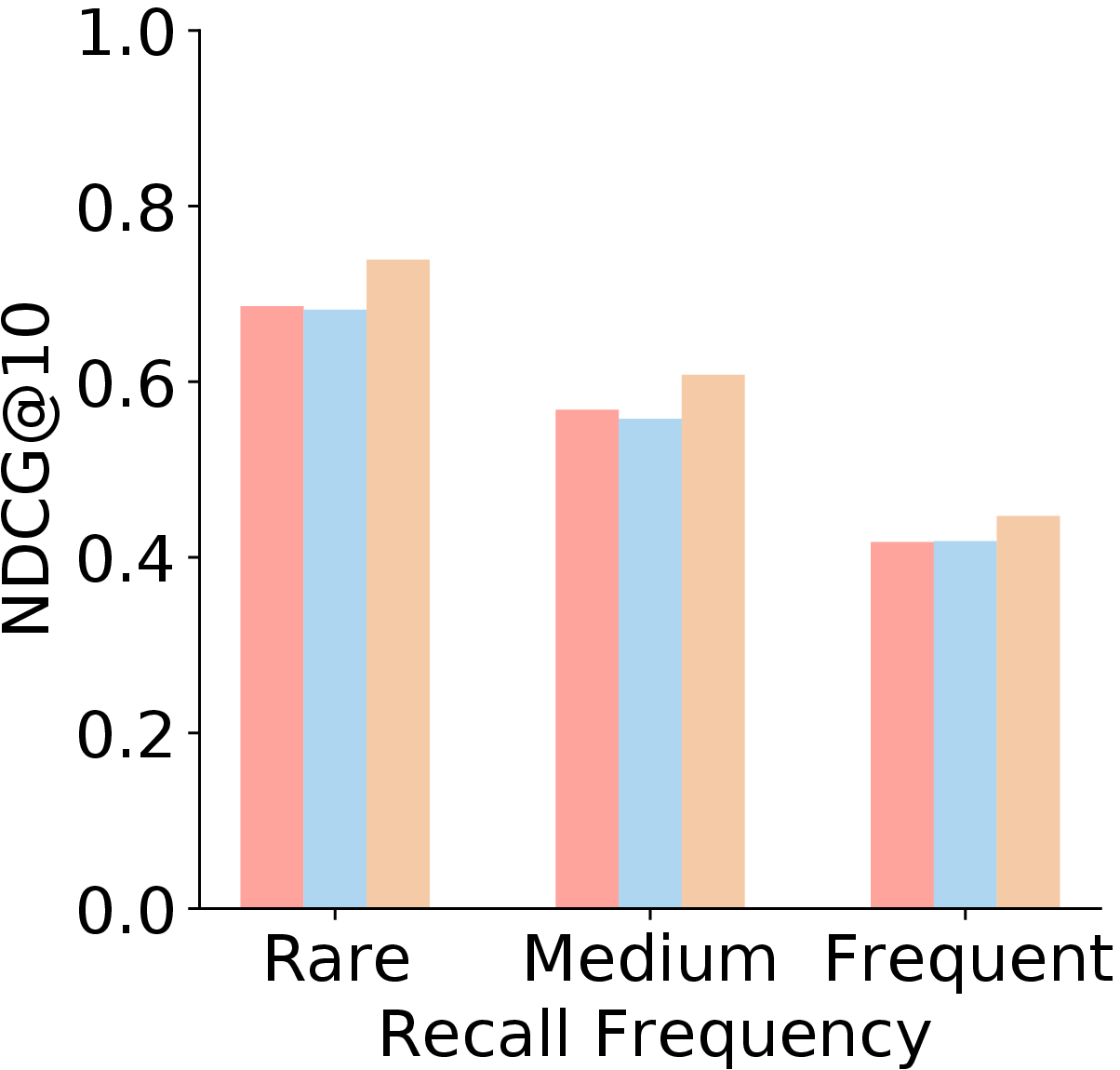}
    } \hfill    
    \subcaptionbox{
        Document Retrieval.\label{fig:longtail_ndcg_b}
    }{
        \includegraphics[width=0.22\textwidth]{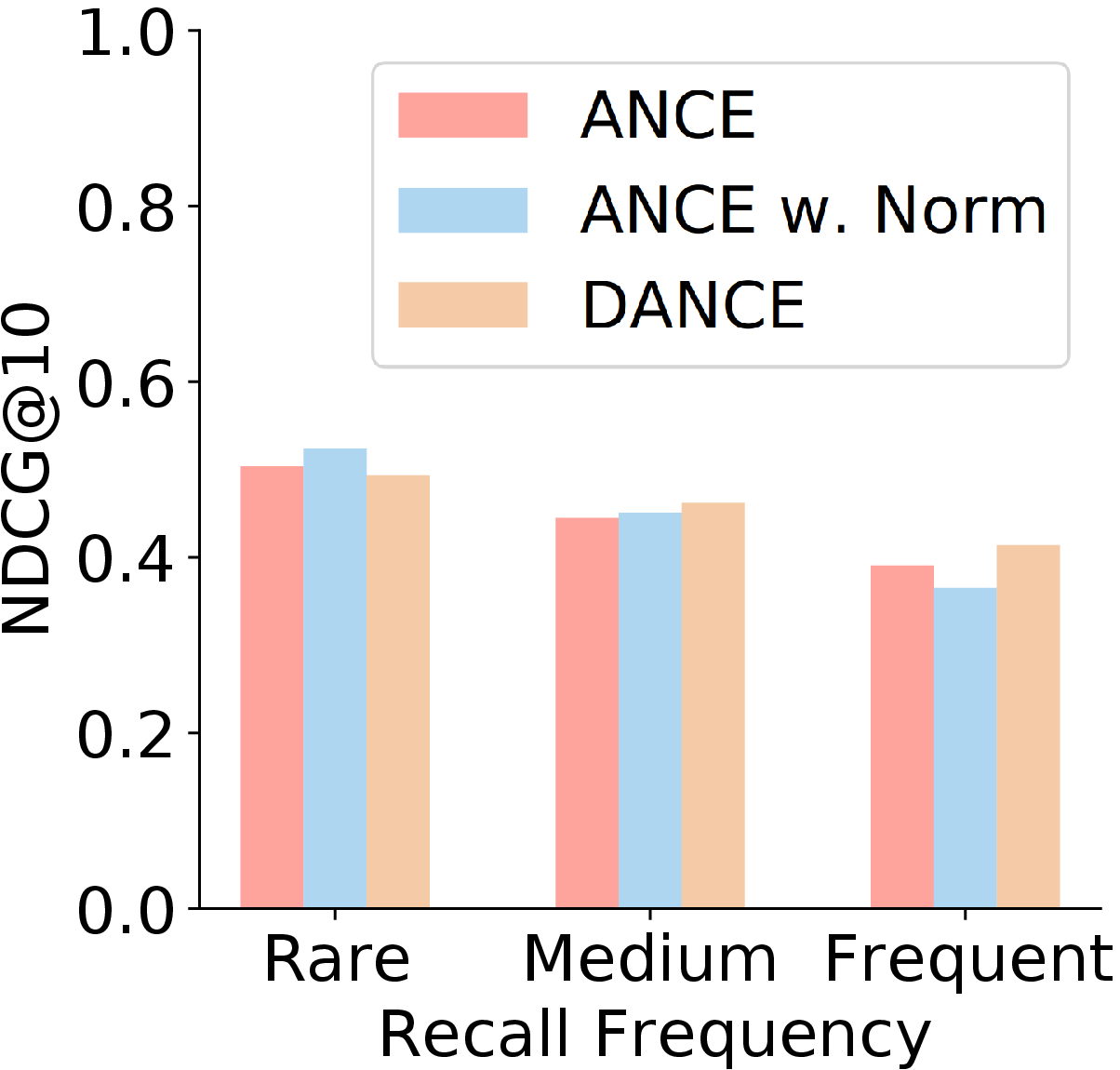}
        
    } \hfill
    \caption{Ranking Performance of Queries and Documents with Different Recall Frequency. The results of query retrieval task and document retrieval task are shown in Figure~\ref{fig:longtail_ndcg_a} and Figure~\ref{fig:longtail_ndcg_b}, respectively. All queries and documents are divided into three groups according to the recall frequency, which includes rare, medium and frequent.}
    \label{fig:longtail_ndcg}
\end{figure}
\begin{figure}[t]
    \centering
    \subcaptionbox{
        Query Retrieval.\label{fig:ndcg_dist_a}
    }{
        \includegraphics[width=0.21\textwidth]{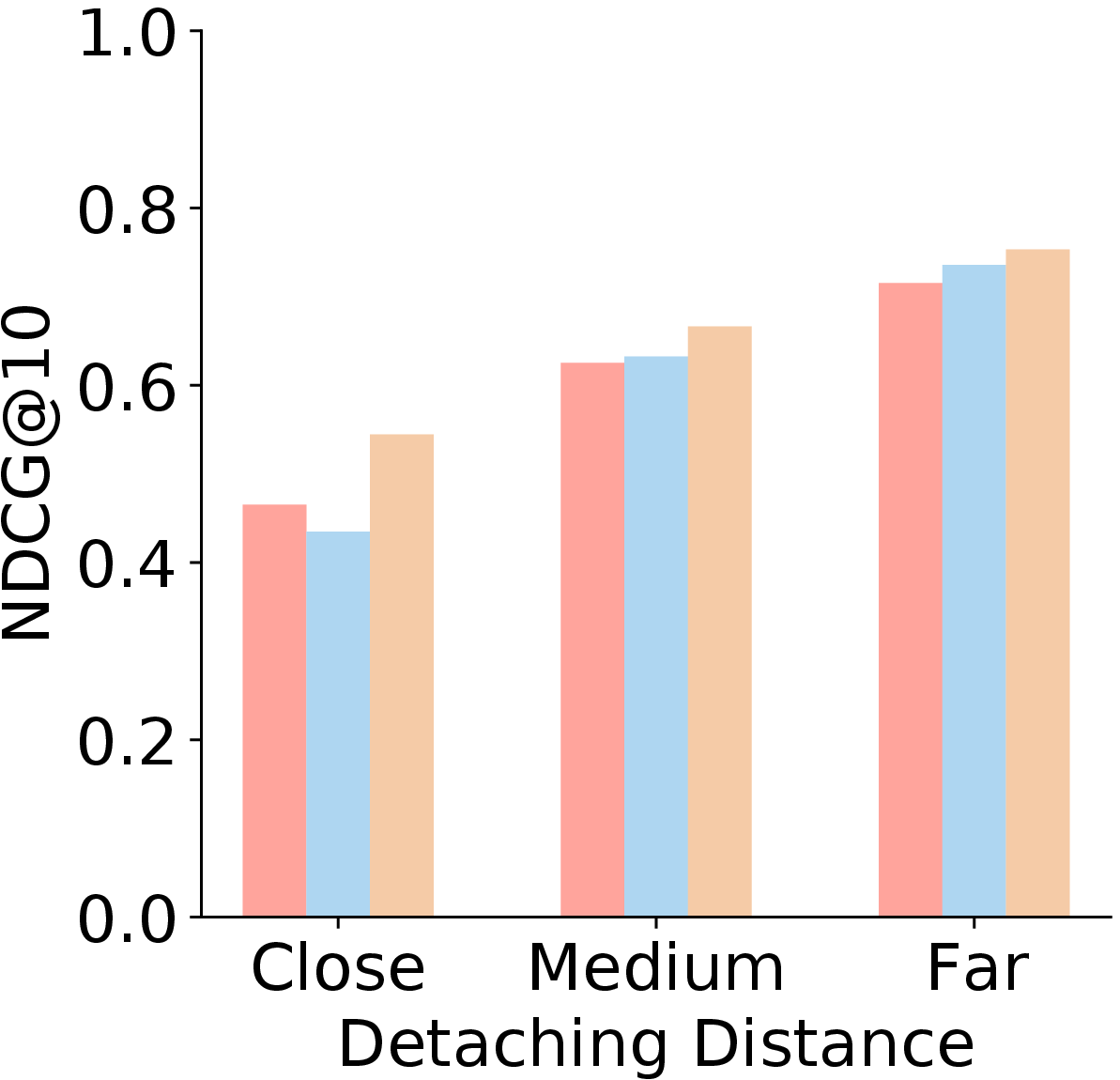}
    } \hfill    
    \subcaptionbox{
        Document Retrieval.\label{fig:ndcg_dist_b}
    }{
        \includegraphics[width=0.21\textwidth]{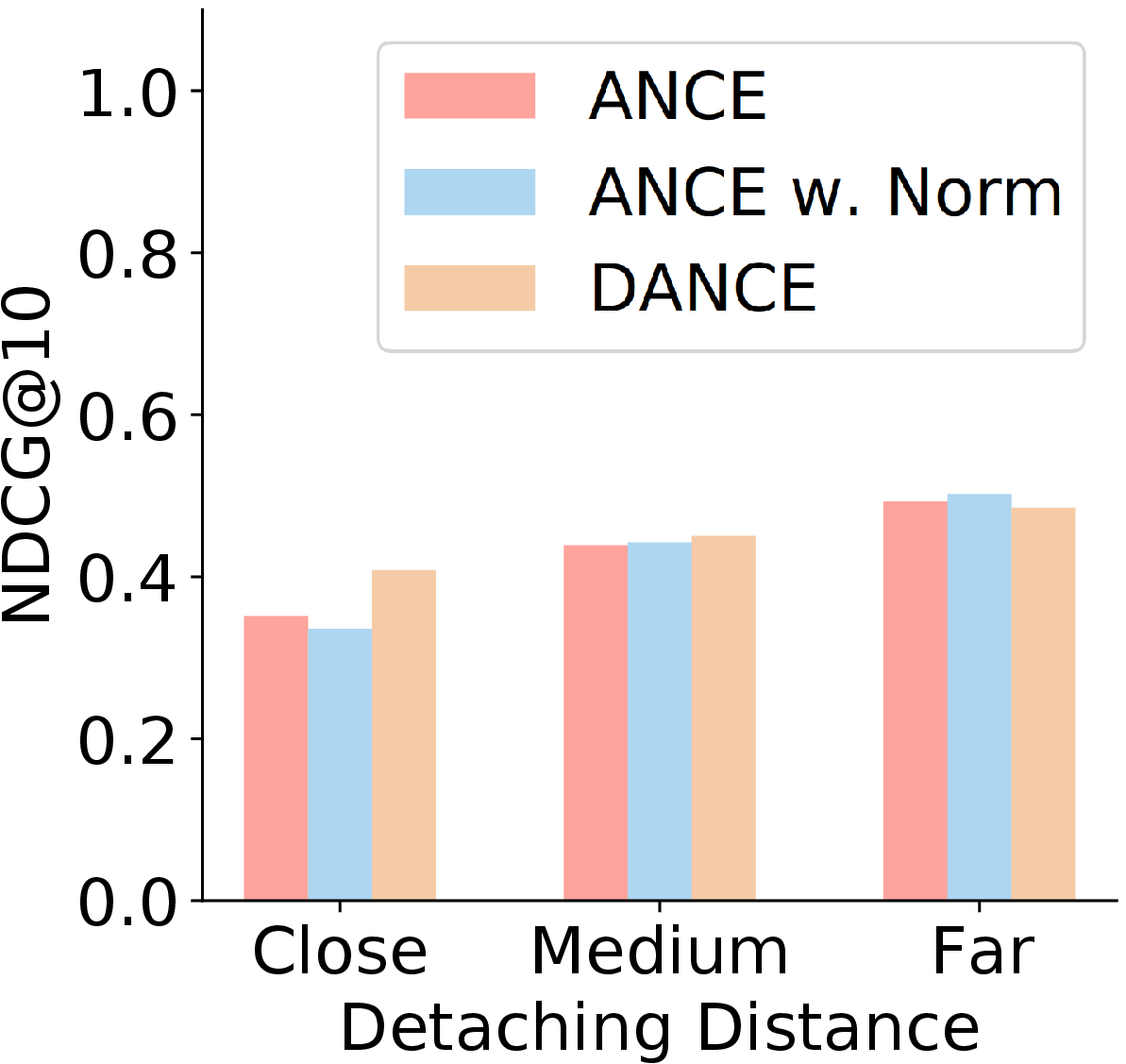}
    } \hfill
    % \newline
    % \subcaptionbox{
    %     Query Retrieval and Doc-Que Distance.\label{fig:ndcg_dist_c}
    % }{
    %     \includegraphics[width=0.225\textwidth]{pictures/x_scaled_D-to-Q-distance_y_d2q-retrieval-NDCG.pdf}
    % } \hfill    
    % \subcaptionbox{
    %     Document Retrieval and Que-Doc Distance.\label{fig:ndcg_dist_d}
    % }{
    %     \includegraphics[width=0.225\textwidth]{pictures/x_scaled_Q-to-D-distance_y_q2d-retrieval-NDCG.pdf}
    % \newline
    % \subcaptionbox{
    %     Doc-Doc Distance.\label{fig:ndcg_dist_c}
    % }{
    %     \includegraphics[width=0.225\textwidth]{pictures/x_D-recall-in-q2d_y_scaled_D-to-D-distance.pdf}
    % } \hfill    
    % \subcaptionbox{
    %     Que-Que Distance.\label{fig:ndcg_dist_d}
    % }{
    %     \includegraphics[width=0.225\textwidth]{pictures/x_Q-recall-in-d2q_y_scaled_Q-to-Q-distance.pdf}
    % } \hfill
    
    \caption{Ranking Performance of Queries and Documents with Different Detaching Distances. The ranking results on query retrieval task and document retrieval task are shown in Figure~\ref{fig:ndcg_dist_a} and Figure~\ref{fig:ndcg_dist_b}, respectively.
    All queries and documents are divided into three groups, close, medium and far, according to the detaching distances. These three groups have almost same numbers of queries or documents. The queries or documents with larger detaching distances are farther away from the query or document group.}
    \label{fig:ndcg_dist}
\end{figure}

\begin{table*}[!ht]
\small
  \caption{Case Studies. Three cases are selected from the development set of MS MARCO to qualitatively analyze the ranking effectiveness of \texttt{DANCE}.}
  \label{tab:case_study_dance_win}
\begin{tabular}{p{0.09\linewidth}|p{0.4\linewidth}|p{0.4\linewidth}}
\hline 
\textbf{Model}  & \textbf{DANCE}& \textbf{ANCE} 
\tabularnewline 
\hline \hline 
\textbf{Query} & \multicolumn{2}{p{0.8\linewidth}}{1101870: willie mays worth}
\tabularnewline 
\hline
% Query Recall &  Medium & Rare
% \tabularnewline 
% \hline
Doc. Retrieval & relevant doc rank 1, recall frequency Medium & relevant doc rank 8, recall frequency Medium
\tabularnewline 
\hline
Que. Retrieval & relevant query rank 2, recall frequency Medium  & relevant query rank 6, recall frequency Rare
\tabularnewline 
\hline
%Top Docs & Rare 54, Medium 40, Frequent 6 &  Rare 51, Medium 44, Frequent 5
%\tabularnewline 
%\hline
% Relevant Doc& rank 1, Medium &  rank 8, Medium
% \tabularnewline 
% \hline
ID, Rank& D1658662, 1 & D1411505, 1
\tabularnewline 
\hline
Title& Who is Willie Mays? Biography, gossip, facts? & Mo Williams Maurice Williams Jr
\tabularnewline 
\hline
Snippet 
&
... Advertisement Willie Howard Mays Jr. (born May 6 1931) is a retired American professional baseball player who spent the majority of his major league career with the New York and San Francisco Giants before finishing with the New York Mets.... 
&
... Williams (born December 19, 1982) is an American professional basketball player who currently plays for the Portland Trail Blazers of the National Basketball Association (NBA) ...
\tabularnewline 
\hline \hline
\textbf{Query} & \multicolumn{2}{p{0.7\linewidth}}{414714: is it still dangerous to go to the place of where the chernobyl happened}
\tabularnewline 
\hline
% Query Recall &  Medium & Rare
% \tabularnewline 
% \hline
Doc. Retrieval & relevant doc rank 2, recall frequency Rare & relevant doc rank 6, recall frequency Rare
\tabularnewline 
\hline
Que. Retrieval & relevant query rank 1, recall frequency Medium  & relevant query rank 2, recall frequency Rare
\tabularnewline 
\hline
% Top Docs & Rare 82, Medium 18, Frequent 0 &  Rare 81, Medium 19, Frequent 0
% \tabularnewline 
% \hline
% Relevant Doc& rank 2, Rare &  rank 6, Rare
% \tabularnewline 
% \hline
ID, Rank& D1761818, 2 & D2092055, 2
\tabularnewline 
\hline
Title& How radiation-safe are short-term trips to the Chernobyl Zone? & .
\tabularnewline 
\hline
Snippet 
&
... Up to now in the Zone there are places with considerably elevated and perhaps even deadly radiation. A prolonged, careless stay at such places can lead to radiation injuries of the body and, perhaps, even to chronic radiation sickness ... 
&
Preface: The Chernobyl Accident On 26 April 1986, the most serious accident in the history of the nuclear industry occurred at Unit 4 of the Chernobyl nuclear power plant in the former Ukrainian Republic of the Soviet Union ...
\tabularnewline 
\hline\hline
% \textbf{Query} & \multicolumn{2}{p{0.8\linewidth}}{1094982: how tall should vent stack be}
% \tabularnewline 
% \hline
% Query Recall & Rare & Medium
% \tabularnewline 
% \hline
% Top Docs & Rare 96, Medium 4, Frequent 0 &  Rare 93, Medium 7, Frequent 0
% \tabularnewline 
% \hline
% Relevant Doc & rank 2, Rare &  rank 4, Rare
% \tabularnewline 
% \hline
% ID, Rank& D2177011, 2 & D1076886, 2
% \tabularnewline 
% \hline
% Title& Thread: DWV stack height & The Difference Between Vent Stacks and Stack Vents
% \tabularnewline 
% \hline
% Snippet 
% &
% ... Re: DWV stack height Bruce According to the IRC the vent should extend at least 6"" above the roof or 6"" above the anticipated snow accumulation. This needs to be reported that the vent needs to be extended accordingly and a properly installed flashing installed ... 
% &
%  ... There are many differences between a vent stack and stack vents. To make sure you’re up to code, our Tampa plumbing experts have outlined these differences and the requirements that are essential to meet with each. Vent Stacks A vent stack does not carry waste and is only a stack for venting ...
% \tabularnewline 
% \hline\hline
\textbf{Query} & \multicolumn{2}{p{0.8\linewidth}}{764139: what is ladder move}
\tabularnewline 
\hline
% Query Recall & Medium & Rare
% \tabularnewline 
% \hline
Doc. Retrieval & relevant doc rank 3, recall frequency Rare & relevant doc rank 10, recall frequency Rare
\tabularnewline 
\hline
Que. Retrieval & relevant query rank 2, recall frequency Medium  & relevant query rank 20, recall frequency Rare
\tabularnewline 
\hline
%Top Docs & Rare 88, Medium 9, Frequent 3 &  Rare 90, Medium 8, Frequent 2
%\tabularnewline 
%\hline
% Relevant Doc & rank 3, Rare &  rank 10, Rare
% \tabularnewline 
% \hline
ID, Rank& D3220966, 3 & D3268840, 3
\tabularnewline 
\hline
Title& Your next career move: The ladder or the lattice? & What Is a Lateral Move in Reference to Employment?
\tabularnewline 
\hline
Snippet 
&
... Up is not the only way forward Climbing the ladder is the traditional model for career growth, taking a single pathway upward through the corporate hierarchy ... 
&
 ... A lateral move in employment is when an employee transfers to a different department in the same company or to a different company -- without any significant change in his salary ...
\tabularnewline 
\hline 
\end{tabular}
\end{table*}

\subsection{Effectiveness in Different Scenarios}
\label{subsec:effect_scenes}

We conduct these experiments to evaluate the ranking performance of \texttt{DANCE} in two testing scenarios, recall frequency and detaching distance. The NDCG@10 scores on both query retrieval task and document retrieval task are shown. Three models, \texttt{ANCE}, \texttt{ANCE w. Norm} and \texttt{DANCE} are evaluated in our experiments.

We first study the ranking performance with different recall frequencies, as shown in Figure~\ref{fig:longtail_ndcg}.
As expect, \texttt{DANCE} shows consistent improvements in the query retrieval task over \texttt{ANCE} and \texttt{ANCE w. Norm} with different recall frequencies, which derives from training with the additional training object, query retrieval task. 
In the document retrieval task, \texttt{DANCE} improves the ranking effectiveness on the queries that are frequently recalled and maintains a comparable performance on the queries of rare and medium frequencies. 
The improvement on frequent queries manifests that \texttt{DANCE} can learn more fine-grained representations for these queries and position them more appropriately in the embedding space. The main reason is that \texttt{DANCE} assigns more balanced optimization frequencies for queries during training with the query retrieval task, which is revealed in the previous experiment (Sec.~\ref{sec:results:recall}).
% Considering the observation in previous experiment (Sec.~\ref{sec:results:recall}), \texttt{DANCE} encourages queries to be recalled more uniformly during contrastive training on the query retrieval task, which sufficiently optimizes query encoders during training and thrives on learning the hard query representations.

In Figure~\ref{fig:ndcg_dist}, we evaluate retrieval effectiveness with different detaching distances.
All three dense retrievers achieve better performance on both query and document retrieval tasks when the embedding distributions of queries and documents move towards detaching direction in the space.
These queries and documents with larger detaching distances are more discriminative and usually positioned in the scattered area of the embedding space, which derives from the contrastive learning that pushes the negative queries and documents away from the related documents and queries, respectively. 
It is noteworthy that \texttt{DANCE} improves the query retrieval performance of the documents with closer distances by most, as shown in Figure~\ref{fig:ndcg_dist_a}. Even though these documents are more confusable, \texttt{DANCE} shows its effectiveness in learning fine-grained query representations and conducting a more effective query embedding space by aligning the related query-document pairs and pushing the unrelated queries away from the given document.
%This suggests that,  \texttt{DANCE} assigns proper positions for confusable document in the dense area to improve the ``alignment''. Thus training on the dual query retrieval task can in turn improve the performance of corresponding queries in the main document retrieval task.
% Thus, the queries and documents with smaller outlier distance are more confusable and locate in a dense area of the embedding space and have higher probabilities to be recalled during the contrastive training. 
% \texttt{DANCE} can achieve better retrieval performance on these queries by learning more fine-grained query representations to better align the queries and related documents and making the documents more discriminative in the embedding space, which helps to better distinguish the actual matched document from the documents closely distributed to the given queries in the embedding space.

\subsection{Case Study}
Finally, we show three cases that are selected from the development set of MS MARCO in Table~\ref{tab:case_study_dance_win} to analyze the ranking effectiveness.

The dense retrievers, such as \texttt{ANCE}, indeed show their effectiveness in conducting semantic matches and dealing with the vocabulary mismatch problem. 
As shown in the second case, the given query asks the safety of the place where the ``chernobyl'' accident happened. \texttt{ANCE} ranks the confusing document describing the general background of ``chernobyl accident'' at a top rank. Such a document is semantically related but is off-topic to the given query. In the other cases, the given queries ask about information of ``Willie May'' and ``ladder move'', but \texttt{ANCE} ranks documents that are about ``Maurice Williams'' and ``lateral move'' with higher ranks. These documents are unrelated and off-topic to the given queries, but \texttt{ANCE} shows less effectiveness to distinguish the actually matched documents from such confusable documents, making the performance of dense retriever worse than matching with discrete bag-of-words.

Different from \texttt{ANCE}, \texttt{DANCE} incorporates an additional query retrieval task training object, which mainly focuses on learning query representations and optimizing the embedding space. Under the contrastive training, the query retrieval task learns query likelihood and keeps the ``uniformity'' and ``alignment'' in the embedding space. It benefits all three cases and helps to achieve better ranking performance on both query retrieval task and document retrieval task compared with \texttt{ANCE}. To sufficiently train the query representations, \texttt{DANCE} increases the recall frequency of these queries in the query retrieval task, making these queries have more probabilities to be sampled during the contrastive training. Our case studies show that \texttt{DANCE} can assign the matched documents with top ranks, manifesting \texttt{DANCE} can better ``align'' the matched query-document pairs from confusable document clusters by learning more fine-grained query representations.

\section{Conclusion}
% 0. contrastive learning for dense retriever.
% 1. dual -> fine tune query embedding. (how dual contributes for document retreival in here)
% 2. visualization: concentrated query embedding -> smooth & uniform. 
% 3. uniform distr -> more query recall times up -> better position ->  retrieval performance
% future work: improve performance contrastive learning.

% DANCE addresses a novel effective training paradigm for the dense retrievers, where richer supervised information is learned from unaltered existing relevance label. With the introduction of representation normalization and dual task training, DANCE improves the alignment of relevant query-document pairs and the uniformity of query embedding space that is less researched in previous studies. Such uniform and regularized representation space further leads to a stable growth in document retrieval performance. Last but not least, we use abundant analytical experiments to provide insights about how the embedding distribution is modified in our method as well as what a well-learned representation space can contribute to the general dense retriever training.

In this paper, we propose a training paradigm, Contrastive Dual Learning for Approximate Nearest Neighbor (\texttt{DANCE}), to train dense retrievers. \texttt{DANCE} introduces the additional training object, query likelihood, in dense retriever training to learn query and document representations. Different from \texttt{ANCE}, \texttt{DANCE} concentrates the query embeddings and assigns more uniform recall frequency to queries to sufficiently optimize their representations, while the document embedding distribution is optimized to be more scattered and discriminative. Through such embedding space optimization, \texttt{DANCE} achieves better ranking performance than the previous state-of-the-art dense retriever \texttt{ANCE}, especially for the documents that are more frequently recalled during contrastive training on the document retrieval task. Even these documents have smaller detaching distances and are hard to distinguish, \texttt{DANCE} shows better performance on them by learning more fine-grained query representations, better aligning related query-document pairs, and forming a more uniform and smooth embedding space for retrieval tasks. 
The observations of our work provide some possible directions to further improve dense retriever effectiveness and sufficiently optimize the embedding space by learning more effective query and document representations during contrastive training.
\section*{Acknowledgments}
This work is supported by the National Key Research and Development Program of China (No. 2020AAA0106501) and  Beijing Academy of Artificial Intelligence (BAAI).
\balance
% \newpage
\bibliographystyle{ACM-Reference-Format}
% \bibliography{reference.bib}
\bibliography{ref_rebib.bib}

\end{document}